\documentstyle{article}

\title{Quantum Number Fractionalization in Antiferromagnets
\thanks{Lectures given by R. B. Laughlin
at the Workshop
with Learning ``Field Theories for Low-dimensional Condensed Matter Systems:
Spin Systems and Strongly Correlated Electrons''
 --  August, 31 - September, 11 1997 --
Chia Laguna, Sardinia, Italy.} }

\author{R. B. Laughlin\\
        Department of Physics, Stanford University\\
        Stanford, California 94305\\
\and
        D. Giuliano\\
        Dipartimento di Scienze Fisiche\\
        Universit\`a ``Federico II'' di Napoli\\
        and I.N.F.N. Sezione di Napoli\\
        Mostra d'Oltremare, Pad. 19\\
        I--80100 Napoli, Italy\\
\and
        R. Caracciolo\\
        Dipartimento di Fisica Teorica\\
        Universit\`a di Torino\\
        and I.N.F.N. Sezione di Torino\\
        Via P. Giuria, 1\\
        I--10125 Torino, Italy\\
\and
        Olivia L. White\\
        Mathematical Institute\\
        University of Oxford\\
        24-29 St. Giles'\\
        Oxford OX1 3LB\\
        U.K.}

\input{epsf}

\begin{document}

\maketitle

\newpage

\tableofcontents

\newpage

\section{Introduction}

In these lectures we shall derive and discuss a set of exact
eigenstates of the Haldane-Shastry \cite{haldane,shastry} model,
a realization of the spin-1/2 Heisenberg chain in which the
quantum-disordered  spin liquid ground state and the neutral, spin-1/2
excitations of such systems are particularly easy to understand.  This
behavior is not unique to the model, and in particular occurs in the
Bethe solution of the near-neighbor Heisenberg chain, where it was
first discovered \cite{bethe,fadeev}, but it is more accessible
in this form. The model also makes the relationship of the spin chain
to the fractional quantum Hall effect transparent \cite{kalmeyer}.
The key results are these:

\begin{figure}
\epsfbox{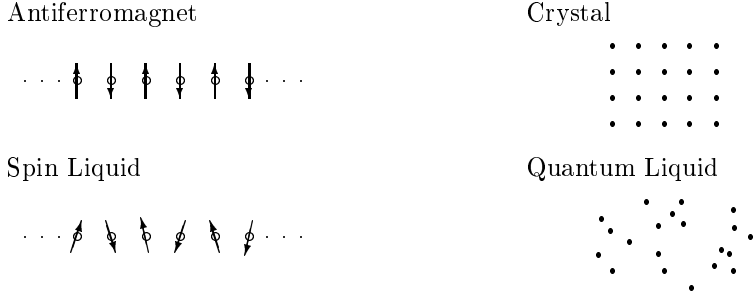}
\caption{Illustration of the quantum spin liquid ground state.}
\end{figure}

\begin{enumerate}

\item The ground state has the same functional form as the fractional
      quantum Hall ground state.

\item This ground state has quantum disorder and has the same relation
      to the antiferromagnetically ordered state that a quantum liquid
      has to a conventional crystal.

\item The elementary excitations of this state are spin-1/2 particles,
      spinons, and not spin waves, which are the elementary excitations
      of an ordered antiferromagnet. Spinons have

      \begin{enumerate}

      \item A relativistic band structure with a Dirac point at
            momentum $\pm \pi/2$.

      \item 1/2 fractional statistics.

      \item $N/2$ allowed momenta rather than $N$.

      \end{enumerate}

\item The ground state is effectively 2-fold degenerate.  The missing
      N/2 states of the spinon are excitations of the other ground
      state.

\item The Hamiltonian has a factorization which may be construed as
       supersymmetric.

\end{enumerate}

\section{Haldane-Shastry Hamiltonian}

Let a lattice of $N$ sites be wrapped onto a unit circle, as
shown in Fig. 2, so that each site may be expressed as a complex
number $z_\alpha$ satisfying

\begin{equation}
z_\alpha^N - 1 = 0
\; \; \; ,
\end{equation}

\noindent
and let each site possess a single unpaired half-integral spin with
its corresponding spin operator $\vec{S}_\alpha$.  The Haldane-Shastry
Hamiltonian is then given by

\begin{equation}
{\cal H}_{HS} = J \left( \frac{2 \pi}{N} \right)^2
\sum_{\alpha < \beta}^N \frac{\vec{S}_\alpha \cdot
\vec{S}_\beta}{ | z_\alpha - z_\beta |^2 }
\; \; \; .
\end{equation}

\noindent
This Hamiltonian is translationally invariant, satisfies

\begin{equation}
[ {\cal H}_{HS} , \vec{S} ] = 0
\; \; \; \; \; \; \; \; \;
\vec{S} = \sum_\alpha^N \vec{S}_\alpha
\; \; \; ,
\end{equation}

\noindent
and possesses the special internal symmetry

\begin{equation}
[ {\cal H}_{HS} , \vec{\Lambda} ] = 0
\; \; \; \; \; \; \; \; \;
\vec{\Lambda} = \frac{i}{2} \sum_{\alpha \neq \beta}
\biggl( \frac{z_\alpha + z_\beta}{z_\alpha - z_\beta} \biggr)
(\vec{S}_\alpha \times \vec{S}_\beta)
\; \; \; .
\label{lambda}
\end{equation}

\bigskip

\begin{center}
{\bf Proof}
\end{center}

\noindent
We proceed by applying the commutation relations

\begin{equation}
[ \vec{S}_{1} , (\vec{S}_{1} \cdot \vec{S}_{2} ) ] = i (\vec{S}_{1}
\times \vec{S}_{2} )
\end{equation}

\begin{equation}
[ (\vec{S}_{1} \times \vec{S}_{2}) , (\vec{S}_{1} \cdot \vec{S}_{2})]
= \frac{i}{2} (\vec{S}_{1} - \vec{S}_{2}) \; \; \; ,
\end{equation}

\noindent
the vector identity

\begin{equation}
\vec{a} \times (\vec{b} \times \vec{c}) = (\vec{a} \cdot \vec{c}) \;
\vec{b} - (\vec{a} \cdot \vec{b}) \; \vec{c}
\; \; \; ,
\end{equation}

\noindent
the fact that

\begin{equation}
\biggl[ \frac{z_{j} + z_{k}}{z_{j} - z_{k}} -  \frac{z_{\ell} +
z_{k}}{z_{\ell} - z_{k}} \biggr] \frac{1}{\mid \! z_{j} - z_{\ell}
\! \mid^2 } = \frac{z_{j} z_{k} z_{\ell}}{(z_{j} - z_{k})
(z_{\ell} - z_{k}) (z_{j} - z_{\ell})} \; \; \; ,
\end{equation}

\noindent
which is totally antisymmetric under permutation of $j$, $k$,
and $\ell$, and

\begin{equation}
\sum_{j \neq k} \frac{z_{j} + z_{k}}{z_{j} - z_{k}} \frac{1}
{\mid \! z_{j} - z_{k} \! \mid^2} = - \sum_{\alpha = 1}^{N-1}
\frac{z_{\alpha}(z_{\alpha} + 1)}{(z_{\alpha} - 1)^3} = 0
\; \; \; .
\end{equation}

\noindent
We have

\begin{displaymath}
\sum_{j \neq k} \sum_{\alpha \neq \beta} \frac{z_{j} + z_{k}}
{z_{j} - z_{k}} \frac{1}{\mid \! z_{\alpha} - z_{\beta} \! \mid^2}
\; [ (\vec{S}_{j} \times \vec{S}_{k}) , (\vec{S}_{\alpha} \cdot
\vec{S}_{\beta})]
\end{displaymath}

\begin{displaymath}
= 4 i \sum_{j \neq k \neq \ell}
\frac{z_{j} + z_{k}}{z_{j} - z_{k}} \frac{1}{ \mid \! z_{j} -
z_{\ell} \! \mid^2 } \biggl[ (\vec{S}_{j} \cdot \vec{S}_{k}) \;
\vec{S}_{\ell} - (\vec{S}_{\ell} \cdot \vec{S}_{k}) \; \vec{S}_{j}
\biggr]
\end{displaymath}

\begin{equation}
+ i \sum_{j \neq k} \frac{z_{j} + z_{k}}{z_{j} - z_{k}}
\frac{1}{ \mid \! z_{j} - z_{k} \! \mid^2 } \; (\vec{S}_{j} -
\vec{S}_{k} ) = 0 \; \; \; . \; \Box
\end{equation}

\bigskip

\noindent
It follows from the commutation relations

\begin{equation}
[ S^a , S^b ] = i \; \epsilon^{abc} S^c
\; \; \; \; \; \; \; \; \; \; \; \; \;
[ S^a , \Lambda^b ] = i \; \epsilon^{abc} \Lambda^c
\; \; \; ,
\label{vector}
\end{equation}

\noindent
that ${\cal H}_{HS}$, $S^2$, and $(\vec{\Lambda} \cdot \vec{S})$
all commute with each other.

\begin{figure}
\epsfbox{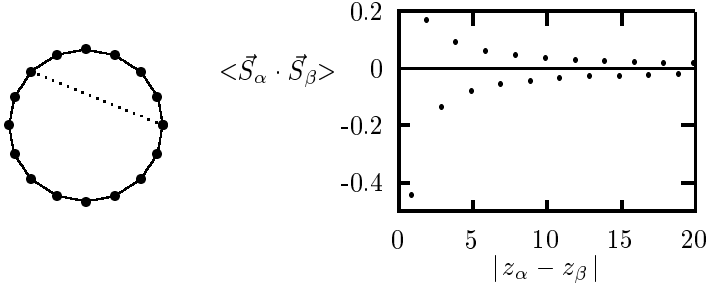}
\caption{Left: Illustration of Haldane-Shastry model.  Right: Magnetic
         correlation function defined by Eq. (28).}
\end{figure}

\section{Ground State}

Let us imagine the spin system to be a 1-dimensional string of boxes
populated by hard-core bosons, the $\downarrow$ spin state
corresponding to an empty box and the $\uparrow$ spin state corresponding
to an occupied one. The ground state wavefunction is a rule by which a
complex number is assigned to each boson configuration.  The total number
of bosons is conserved, as it is physically the same thing as the
eigenvalue of $S^z$.  Let $N$ be even and let $z_1 , ... , z_{N/2}$
denote the locations of $N/2$ bosons, the number appropriate for a
spin singlet. Then the Haldane-Shastry ground state is

\begin{equation}
\Psi (z_1, ... , z_{N/2}) =
\prod_{j<k}^{N/2} (z_j - z_k)^2
\prod_{j=1}^{N/2} z_j
\; \; \; .
\end{equation}

\noindent
Its energy is

\begin{equation}
{\cal H}_{HS} | \Psi \! > = -J \left( \frac{\pi^2}{24} \right)
\left(N + \frac{5}{N} \right) | \Psi \! > \; \; \; .
\end{equation}

\bigskip

\begin{center}
{\bf Proof}
\end{center}

\noindent
We begin by observing that $[S_\alpha^+ S_\beta^- \Psi ]
(z_1 , ... , z_{N/2})$ is identically zero unless one of the arguments
$z_1 , ... , z_{N/2}$ equals $z_\alpha$. We have

\begin{displaymath}
[ \biggl\{ \sum_{\beta \neq \alpha}^{N} \frac{ S_{\alpha}^{+}
S_{\beta}^{-}}{\mid \! z_{\alpha} - z_{\beta} \! \mid^2 }
\biggr\} \Psi] (z_{1}, \ldots , z_{N/2})
\end{displaymath}

\begin{displaymath}
= \sum_{j=1}^{N/2} \sum_{\beta \neq j}^{N} \frac{1}{\mid \! z_j
- z_{\beta} \! \mid^2 } \Psi (z_{1}, \ldots , z_{j-1}, z_{\beta} ,
z_{j+1}, \ldots , z_{N/2})
\end{displaymath}

\begin{displaymath}
= \sum_{j=1}^{N/2} \sum_{\ell = 0}^{N-2} \biggl\{
\sum_{\beta \neq j}^{N} \frac{ z_{\beta} (z_{\beta} - z_{j})^{\ell}}
{ \ell ! \mid \! z_{j} - z_{\beta} \! \mid^2} \biggr\}
(\frac{\partial}{\partial z_{j}})^{\ell} \biggl\{ \Psi
(z_{1}, \ldots , z_{N/2})/z_{j} \biggr\}
\end{displaymath}

\begin{displaymath}
= \sum_{j=1}^{N/2} \biggl\{ \frac{(N-1)(N-5)}{12} z_{j} - \frac{N-3}{2}
 z_{j}^2 \frac{\partial}{\partial z_{j}}
\end{displaymath}

\begin{displaymath}
 + \frac{1}{2} z_{j}^3
 \frac{\partial^2} {\partial z_{j}^2} \biggr\} \biggl\{ \Psi (z_{1},
 \ldots , z_{N/2})/z_{j} \biggr\}
\end{displaymath}

\begin{displaymath}
= \biggl\{ \frac{N(N-1)(N-5)}{24} - \frac{N-3}{2}
\sum_{j \neq k}^{N/2} \frac{2 z_{j}}{z_{j} - z_{k}}
\end{displaymath}

\begin{displaymath}
+ \frac{1}{2} \biggl[ \sum_{j \neq k \neq m}^{N/2} \frac{4 z_{j}^2}
{(z_{j} - z_{k})(z_{j} - z_{m})}
+ \sum_{j \neq k}^{N/2} \frac{2 z_{j}^2}
{(z_{j} - z_{k})^2} \biggr] \biggr\} \Psi (z_{1}, \ldots ,z_{N/2})
\end{displaymath}

\begin{displaymath}
= \biggl\{ \frac{N(N-1)(N-5)}{24} - \frac{N(N-3)}{4}
( \frac{N}{2} - 1) + \frac{1}{2} \biggl[ \frac{2N}{3}
(\frac{N}{2} - 1)(\frac{N}{2} - 2)
\end{displaymath}

\begin{displaymath}
+ \frac{N}{2}(\frac{N}{2} - 1) - \sum_{j \neq k}^{N/2}
\frac{2}{\mid \! z_{j} - z_{k} \! \mid^2 } \biggr] \biggr\}
\Psi (z_{1}, \ldots , z_{N/2})
\end{displaymath}

\begin{equation}
= \biggl\{ - \frac{N}{8} - \sum_{j \neq k}^{N/2}
\frac{1}{\mid \! z_{j} - z_{k} \! \mid^2 } \biggr\}
\Psi (z_{1}, \ldots , z_{N/2}) \; \; \; .
\end{equation}

\noindent
Here we have used the fact that

\begin{equation}
\frac{1}{\mid \! z_{\alpha} - z_{\beta} \! \mid^2} = - \frac{z_{\alpha}
z_{\beta}}{(z_{\alpha} - z_{\beta})^2} \; \; \; ,
\end{equation}

\noindent
i.e. that the Hamiltonian is effectively analytic in the spin
coordinates, and the sums (with sites labeled so that
$z_{N} = 1$)

\begin{equation}
\sum_{\alpha = 1}^{N-1} \frac{1}{\mid \! z_{\alpha} - 1 \! \mid^2}
= \frac{N^2 - 1}{12} \; \; \; ,
\end{equation}

\begin{equation}
\sum_{\alpha = 1}^{N-1} \frac{z_{\alpha}^2}{(z_{\alpha} - 1)^2}
= - \frac{(N-1)(N-5)}{12} \; \; \; ,
\end{equation}

\begin{equation}
\sum_{\alpha = 1}^{N-1} \frac{z_{\alpha}^2}{(z_{\alpha} - 1)}
= \frac{N-3}{2} \; \; \; ,
\end{equation}

\begin{equation}
\sum_{\alpha = 1}^{N-1} z_{\alpha}^2 = -1 \; \; \; ,
\end{equation}

\begin{equation}
\sum_{\alpha = 1}^{N-1} z_{\alpha}^2 (z_{\alpha}-1)
= \ldots =
\sum_{\alpha = 1}^{N-1} z_{\alpha}^2 (z_{\alpha}-1)^{N-3} = 0
\; \; \; ,
\end{equation}

\begin{equation}
\frac{z_{\alpha}^2}{(z_{\alpha}-z_{\beta})(z_{\alpha}-z_{\gamma})} +
\frac{z_{\beta}^2}{(z_{\beta}-z_{\alpha})(z_{\beta}-z_{\gamma})} +
\frac{z_{\gamma}^2}{(z_{\gamma}-z_{\alpha})(z_{\gamma}-z_{\beta})} =
1 \; ,
\end{equation}

\noindent
which are worked out in Appendix A. We also have

\begin{displaymath}
[ \biggl\{ \sum_{\beta \neq \alpha}^{N} \frac{ S_{\alpha}^{z}
S_{\beta}^{z}}{\mid \! z_{\alpha} - z_{\beta} \! \mid^2 }
\biggr\} \Psi] (z_{1}, \ldots , z_{N/2})
\end{displaymath}

\begin{equation}
= \biggl\{ - \frac{N(N^2 - 1)}{48}
+ \sum_{j \neq k}^{N/2} \frac{1}{\mid \! z_{j} - z_{k} \!
\mid^2 } \biggr\} \Psi (z_{1}, \ldots , z_{N/2})
\; \; \; .
\end{equation}

\noindent
This completes the proof, since

\begin{equation}
{\cal H}_{HS} = \frac{1}{2} J (\frac{2\pi}{N})^2 \biggl\{ \sum_{\alpha
\neq \beta} \frac{ S^{+}_{\alpha} S^{-}_{\beta}}{ \mid \! z_{\alpha} -
z_{\beta} \! \mid^2 } + \sum_{\alpha \neq \beta} \frac{ S^{z}_{\alpha}
S^{z}_{\beta}}{ \mid \! z_{\alpha} - z_{\beta} \! \mid^2 } \biggr\}
\; \; \; . \; \; \Box
\end{equation}

\bigskip

\begin{center}
{\bf Important Properties}
\end{center}

\begin{enumerate}

\item {\bf Reality:} Since $z_{\alpha}$ lies on the unit circle we
      have

      \begin{displaymath}
      \Psi^{*}(z_{1} , \ldots , z_{N/2}) = \prod_{j < k}^{N/2}
      (z^{*}_{j} - z^{*}_{k})^2 \prod_{j}^{N/2} z^{*}_{j}
      \end{displaymath}

      \begin{equation}
      = \prod_{j < k}^{N/2}
      (z_{k} - z_{j})^2 \prod_{j}^{N/2} z_{j}^{1-N} =
      \Psi (z_{1} , \ldots , z_{N/2})
      \; \; \; .
      \end{equation}

      Thus $\Psi$ is real despite being
      a polynomial in the complex variables $z_{j}$.

\item {\bf Translational Invariance:}  $\Psi$ is translated one lattice
      spacing by multiplying each of its arguments by $z = \exp (i2\pi
      / N)$.  Since it is a homogeneous polynomial of degree
      $(N/2)^2$ we have

      \begin{equation}
      \Psi (z_{1} z , \ldots , z_{N/2} z ) = \exp (iN\pi / 2) \;
      \Psi (z_{1} , \ldots , z_{N/2}) \; \; \; .
      \end{equation}

      The crystal momentum of the state, i.e. the phase it
      acquires under translation, is thus 0 or $\pi$, depending on
      the value of the even integer $N$.

\item {\bf Spin Rotational Invariance:}  To prove $\Psi$ is a spin
      singlet it suffices to show that it is an eigenstate of $S^z$
      with eigenvalue zero and that it is destroyed by the spin
      lowering operator $S^{-}$.  The former is true for any
      wavefunction in which the number of $z_{j}$ is constrained
      to $N/2$. For the latter we have

      \begin{displaymath}
      [S^{-}\Psi](z_{2}, \ldots ,z_{N/2}) = \sum_{\alpha = 1}^{N}
      \Psi (z_{\alpha},z_{2}, \ldots ,z_{N/2})
      \end{displaymath}

      \begin{equation}
      = \lim_{z_1 \rightarrow 0} \;
      \sum_{\ell = 1}^{N-1} \frac{1}{\ell !}\biggl\{
      \sum_{\alpha = 1}^{N} z_{\alpha}^{\ell} \biggr\}
      \frac{\partial^{\ell}}{\partial z_{1}^{\ell}} \Psi
      (z_{1},z_{2}, \ldots ,z_{N/2})  = 0
      \; \; \; ,
      \label{singlet}
      \end{equation}

      since

      \begin{equation}
      \sum_{\alpha = 1}^{N} z_{\alpha}^{\ell} = N \; \delta_{\ell 0}
      \pmod{N} \; \; \; .
      \end{equation}

      This implies that the wavefunction is the same with the roles
      of $\uparrow$ and $\downarrow$ reversed or, more generally,
      with the quantization axis taken to be an arbitrary direction
      in spin space.

\item {\bf Quantum Disorder:} In Fig. 2 we plot the spin-spin
      correlation function

      \begin{displaymath}
      < \! \vec{S}_\alpha \cdot \vec{S}_\beta \! >
      \end{displaymath}

      \begin{equation}
      = \frac{3}{2}
      (\frac{N}{2} - 1) \frac{ \sum_{z_3 , ... , z_{N/2}} |
      \Psi (z_\alpha , z_\beta , z_3 , ... , z_{N/2}) |^2 }
      { \sum_{z_2 , ... , z_{N/2}} | \Psi (z_\alpha , z_2 , ... ,
      z_{N/2}) |^2 } - \frac{3}{4} \; \; \; .
      \label{corr}
      \end{equation}

      The convergence of this function to zero as $|z_\alpha - z_\beta|
      \rightarrow \infty$ shows that $| \Psi \! >$ has no long-range
      order and is a spin liquid. The fall-off is slow, however, and
      this is important. Strongly-disordered spin liquids, i.e. with
      exponential decay of correlations on a length scale $\xi$, are
      easy to construct when the spin in the unit cell is integral.
      They have an energy gap $\Delta = \hbar v / \xi$, where $v$ is
      the spin-wave velocity of a nearby ordered state. But
      strongly-disordered spin liquids {\it cannot} be stabilized
      with short-range interactions when the spin in the unit cell is
      half-integral.  The excitation spectrum in this case is always
      gapless \cite{haldane2}.

\item {\bf Factorizability:} The fact that $| \Psi \! >$ is a product
      of pair factors makes a number of its properties easy to calculate
      by semi-classical monte-carlo techniques.  Let us, for example,
      consider Eq. (\ref{corr}).   Writing

      \begin{displaymath}
      | \Psi (z_1 , ... , z_{N/2}) |^2 = \exp \biggl[ - \phi (z_1 ,
      ... , z_{N/2}) \biggr]
      \end{displaymath}

      \begin{equation}
      \phi (z_1 , ... , z_{N/2}) = - 4 \sum_{j < k} \ln |z_j - z_k |
      \; \; \; ,
      \end{equation}

      we see that the summand is the Boltzmann factor of an equivalent
      finite-temperature classical lattice gas, and that we are
      computing the joint probability for two of these particles to
      reside at sites $z_\alpha$ and $z_\beta$.  This may be done by
      generating a time sequence of configurations using a rule that
      obeys detailed balance, and then simply counting how many times
      the sites $z_\alpha$ and $z_\beta$ are simultaneously occupied.
      The simplest such algorithm is the following.  Let the current
      configuration be $z_1 , ... , z_{N/2}$.

      \begin{enumerate}

      \item Loop on particles $j$.

      \item For this particle, roll the dice to choose a direction.
            Compute

            \begin{equation}
            z_j ' = z_j \exp ( \pm i 2\pi / N )
            \end{equation}

            depending on the outcome.

      \item Compute

            \begin{equation}
            f = \prod_{k \neq j} \biggl| \frac{ z_j ' - z_k }
            {z_j - z_k} \biggr|^4 \; \; \; .
            \end{equation}

      \item If $f > 1$ update $z_j$ to $z_j '$.

      \item If $f < 1$, roll the dice to generate a real number $x$
            between 0 and 1.  Update $z_j$ to $z_j '$ if $x < f$ but do
            nothing otherwise.

      \end{enumerate}

      Fig. 2 was generated using this algorithm.

\item {\bf Degeneracy:} The Haldane-Shastry ground state is not
      degenerate, but it is nearly so. The alternate ground state is

      \begin{equation}
      \Psi ' (z_1, ... , z_{N/2}) =
      \prod_{j<k}^{N/2} (z_j - z_k)^2
      \biggl[ 1 - \prod_{j=1}^{N/2} z_j^2 \biggr] \; \; \; .
      \label{alternate}
      \end{equation}

      It has crystal momentum $\pi$ greater than that of $\Psi$ and has
      energy

      \begin{equation}
      {\cal H}_{HS} | \Psi ' \! > = -J \left( \frac{\pi^2}{24} \right)
      \left(N - \frac{7}{N} \right) | \Psi ' \! > \; \; \; .
      \end{equation}

      It is equivalent to the original vacuum plus a pair of spinons
      excited out of the vacuum into a singlet with total momentum
      $\pi$.

\end{enumerate}

\section{Spinons}

\begin{figure}
\epsfbox{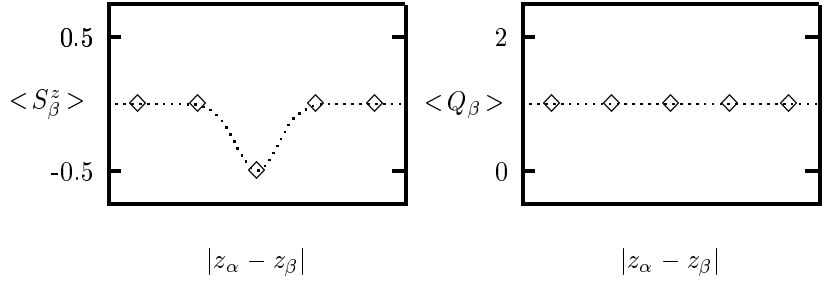}
\caption{Spin and charge profiles of the localized spinon
         $|\Psi_\alpha \! >$ defined by Eq. (34).  The dotted lines
         are a guide to the eye.}
\end{figure}

Let the number of sites $N$ be odd and let

\begin{equation}
\Psi_{\alpha} (z_{1} , ... , z_{M}) =
\prod_{j}^{M} (z_{\alpha} - z_{j}) \;
\prod_{j < k}^{M} (z_{j} - z_{k})^2
\prod_{j}^{M} z_{j} \; \; \; ,
\label{spinon1}
\end{equation}

\noindent
where $M = (N - 1)/2$. This is a $\downarrow$ spin on site $\alpha$
surrounded by an otherwise featureless singlet sea.  We have

\begin{equation}
\sum_{\beta \neq \alpha}^N S_{\beta}^{-} \Psi_{\alpha} = 0
\; \; \; ,
\label{singlet2}
\end{equation}

\noindent
per Eq. (\ref{singlet}). The combination of these states given by

\begin{equation}
\Psi_{m} (z_{1} , \ldots , z_{M}) = \sum_\alpha^N \;
(z_{\alpha}^{*})^{m} \; \Psi_{\alpha} (z_{1} , \ldots , z_{M})
\; \; \; ,
\label{spinon}
\end{equation}

\noindent
with $0 \leq m \leq (N-1)/2$, is an exact eigenstate of the
Hamiltonian with eigenvalue

\begin{equation}
{\cal H}_{HS} | \Psi_{m} \! > = \biggl\{ - J (\frac{\pi^2}{24}) ( N -
\frac{1}{N}) + \frac{J}{2} (\frac{2\pi}{N})^2 m(\frac{N-1}{2} - m)
\biggr\} | \Psi_{m} \! > \; \; \; .
\end{equation}

\bigskip

\begin{center}
{\bf Proof}
\end{center}

\noindent
Following Haldane \cite{haldane} we consider a wavefunction of the
general form

\begin{equation}
\Psi (z_{1} , \ldots , z_{M}) = \Phi (z_{1}, \ldots ,z_{M}) \;
\prod_{j < k}^{M} (z_{j} - z_{k})^2 \prod_{j}^{M} z_{j}
\; \; \; ,
\end{equation}

\noindent
where $M = (N-1)/2$ and $\Phi$ is a homogeneous symmetric polynomial of
degree less than $N - 2M + 2$.  This latter condition causes $\Psi$ to
be a polynomial of degree less than $N+1$ in each of its variables
$z_{j}$, and thus allows the Taylor expansion technique used for the
ground state to be applied.  Doing so, we find that

\begin{equation}
{\cal H}_{HS} \Psi = \frac{J}{2} (\frac{2\pi}{N})^2
\biggl\{ \lambda +\frac{N}{48}(N^2 - 1) + \frac{M}{6}
(4 M^2 - 1) - \frac{N}{2} M^2 \biggr\} \; \Psi \; \;\; ,
\end{equation}

\noindent
provided that $\Phi$ satisfies

\begin{equation}
\frac{1}{2} \biggl\{ \sum_{j}^{M} z_{j}^2 \frac{\partial^2
\Phi} {\partial z_{j}^2} + \sum_{j \neq k}^M \frac{4
z_{j}^2} {z_{j}-z_{k}} \frac{\partial \Phi}{\partial z_{j}}
\biggr\}
- \frac{N-3}{2} \sum_{j}^{M} z_{j} \frac{\partial \Phi}
{\partial z_{j}} = \lambda \Phi \; \; \; .
\label{eig2}
\end{equation}

\noindent
Let us now consider the polynomial

\begin{equation}
\Phi_{A}(z_{1} , \ldots , z_{M}) = \prod_{j}^{M} (z_{A} - z_{j})
= \sum_{m=0}^{M} z_{A}^{m} P_{m} (z_{1}, \ldots ,z_{M}) \; \; \; ,
\end{equation}

\noindent
where $z_{A}$ is not necessarily a lattice site.  We have

\begin{displaymath}
\frac{1}{2} \biggl\{ \sum_{j}^{M} z_{j}^2 \frac{\partial^2
\Phi_{A}} {\partial z_{j}^2} + \sum_{j \neq k}^M \frac{4
z_{j}^2} {z_{j}-z_{k}} \frac{\partial \Phi_{A}}{\partial z_{j}}
\biggr\}
- \frac{N-3}{2} \sum_{j}^{M} z_{j} \frac{\partial \Phi_{A}}
{\partial z_{j}}
\end{displaymath}

\begin{displaymath}
= 2 \sum_{j < k} \biggl[ \frac{z_{j}^2}{(z_{j} - z_{A})
(z_{j} - z_{k})} + \frac{z_{k}^2}{(z_{k} - z_{A}) (z_{k} - z_{j})}
\biggr] \Phi_{A}
\end{displaymath}

\begin{displaymath}
- \frac{N-3}{2} \sum_{j}^{M} \frac{z_{j}}{z_{j} - z_{A}}
\Phi_{A}
\end{displaymath}

\begin{displaymath}
= \biggl\{ M(M-1) - z_{A}^2 \frac{\partial^2}{\partial z_{A}^2}
- \frac{N-3}{2} \biggl[ M - z_{A} \frac{\partial}{\partial z_{A}}
\biggr] \biggr\} \Phi_{A}
\end{displaymath}

\begin{equation}
= \sum_{m=0}^{M} m (\frac{N-1}{2} - m) \; z_{A}^{m} P_{m} \; \; \; .
\end{equation}

\noindent
The proof is completed by multiplying both sides of this equation by
$(z_A^*)^n$ and then summing on lattice sites $z_A$. $\Box$

\bigskip

The state $| \Psi_m \! >$ is a propagating $\downarrow$ spinon with
crystal momentum

\begin{figure}
\epsfbox{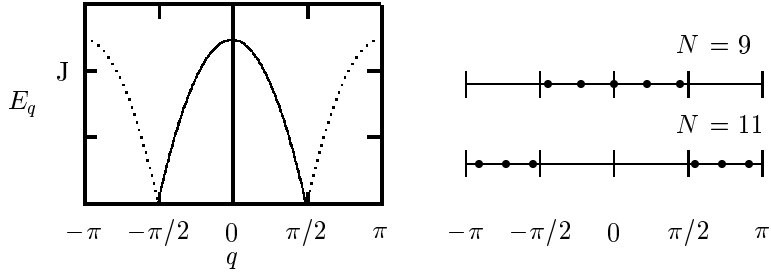}
\caption{Left: Spinon dispersion given by Eq. (46). Right: Allowed
        values of $q$ for adjacent odd $N$.}
\end{figure}

\begin{equation}
q = \frac{\pi}{2} N - \frac{2\pi}{N}(m + \frac{1}{4}) \pmod{2\pi}
\; \; \; ,
\label{momentum}
\end{equation}

\noindent
per the definition

\begin{equation}
\Psi_{m} (z_{1} z , \ldots , z_{M} z) = \exp (iq) \;
\Psi_{m} (z_{1} , \ldots , z_{M}) \; \; \; .
\label{crys2}
\end{equation}

\noindent
Rewriting the eigenvalue as

\begin{equation}
{\cal H} | \Psi_{m} \! > = \biggl\{ - J (\frac{\pi^2}{24}) ( N +
\frac{5}{N} - \frac{3}{N^2}) + E_{q} \biggr\} | \Psi_{m} \! >
\; \; \; ,
\end{equation}

\noindent
we obtain the dispersion relation

\begin{equation}
E_{q} = \frac{J}{2} \biggl[ (\frac{\pi}{2})^2 - q^2 \biggr] \pmod{\pi}
\label{dispersion}
\end{equation}

\noindent
plotted in Fig. 4.  Note that the momenta available to the
spinon span only the inner or outer half of the Brillouin zone,
depending on whether $N - 1$ is divisible by 4.  The spinon
dispersion at low energies is linear in $q$ with a velocity

\begin{equation}
v_{spinon} = \frac{\pi}{2} \frac{Jb}{\hbar} \; \; \; ,
\end{equation}

\noindent
where $b = 2\pi / N$ is the bond length.

The existence of spinons is an automatic consequence of quantum
disorder whenever the total spin in the unit cell is half-integral.
The spin liquid ground state must be a singlet because otherwise
it is ferromagnetic. A singlet is possible only if the total number
of lattice sites $N$ is even.  If $N$ is odd then the total spin can be
no less than 1/2.  But since there can be no physical difference
between even and odd in the limit of large $N$, the system must have
had a neutral spin-1/2 excitation to begin with.

The ground state of the odd-N spin chain is 4-fold degenerate and is
given by $| \Psi_m \! >$ for $m = 0$ and $(N-1)/2$ and their $\uparrow$
counterparts.  This corresponds physically to a ``left-over'' spinon
with momentum $\pm \pi$.

Spinons maintain their identity when more than one of them is
present.  When $N$ is even, for example, the states

\begin{equation}
| \Psi_{mn} \! > = \sum_{\alpha = 1}^{N}  \sum_{\beta = 1}^{N}
\; f_{mn}(z_{\alpha}^* z_{\beta}) \; (z_{\alpha}^{*})^{m}
(z_{\beta}^{*})^{n} | \Psi_{\alpha \beta} \! >
\; \; \; ,
\label{triplet}
\end{equation}

\noindent
where

\begin{equation}
\Psi_{\alpha \beta} (z_{1} , \ldots , z_{M}) =
\prod_{j}^{M} (z_{\alpha} - z_{j}) (z_{\beta} - z_{j}) \;
\prod_{j < k}^{M} (z_{j} - z_{k})^2
\prod_{j}^{M} z_{j} \; \; \; ,
\end{equation}

\noindent
with $M = N/2 - 1$ and

\begin{displaymath}
f_{mn}(z) = \sum_{\ell =0}^{N/2} \; a_{\ell} z^{\ell} - \frac{1}{2}
a_{N/2} z^{N/2}
\end{displaymath}

\begin{equation}
a_{\ell} = \frac{m - n + 2 \ell}{2\ell [ \ell + m - n - 1/2]}
\sum_{k=0}^{\ell -1} \; a_{k}
\; \; \; \; \; \; \; \; \; \; (a_0 = 1)
\label{fmn}
\end{equation}

\noindent
with $m \geq n$, are eigenstates of the Hamiltonian with eigenvalue

\begin{displaymath}
{\cal H}_{HS} | \Psi_{mn} \! > = \biggl\{ - J (\frac{\pi^2}{24}) ( N -
\frac{19}{N} + \frac{24}{N^2}) + \frac{J}{2}(\frac{2 \pi}{N})^2
\end{displaymath}

\begin{equation}
\times \biggl[ m(\frac{N}{2} - 1 - m) + n(\frac{N}{2} - 1 - n) -
\biggl| \frac{m-n}{2} \biggr| \; \biggr] \biggr\} | \Psi_{mn} \! >
\; \; \; .
\end{equation}

\bigskip

\begin{center}
{\bf Proof}
\end{center}

\noindent
Following the procedure we use for one spinon we take $\Phi$ to be
a superposition of states of the form

\begin{displaymath}
\Phi_{A B} = \prod_{j}^{M} (z_{A} - z_{j})(z_{B} - z_{j})
\end{displaymath}

\begin{equation}
= \sum_{m=0}^{M} \; \sum_{n=0}^{M} \; z_{A}^m z_{B}^n \;
P_{m}(z_{1} , \ldots , z_{M}) \;
P_{n}(z_{1} , \ldots , z_{M})
\; \; \; ,
\end{equation}

\noindent
where $z_{A}$ and $z_{B}$ are not necessarily lattice sites. We find
that

\begin{displaymath}
\frac{1}{2} \biggl\{ \sum_{j}^{M} z_{j}^2 \frac{\partial^2
\Phi_{A B}} {\partial z_{j}^2} + \sum_{j \neq k}^M
\frac{4 z_{j}^2} {z_{j}-z_{k}} \frac{\partial \Phi_{A B}}
{\partial z_{j}} \biggr\} - \frac{N-3}{2} \sum_{j}^{M} z_{j}
\frac{\partial \Phi_{A B}}
{\partial z_{j}}
\end{displaymath}

\begin{displaymath}
= \biggl\{ - \frac{z_{A}^2}{z_{A} - z_{B}}
\frac{\partial}{\partial z_{A}}
- \frac{z_{B}^2}{z_{B} - z_{A}}
\frac{\partial}{\partial z_{B}}
- z_{A}^{2} \frac{\partial^2}{\partial z_{A}^2}
- z_{B}^{2} \frac{\partial^2}{\partial z_{B}^2}
\end{displaymath}

\begin{displaymath}
+ (\frac{N-3}{2}) \biggl[
z_{A} \frac{\partial}{\partial z_{A}}
+ z_{B} \frac{\partial}{\partial z_{B}} \biggr]
+ \biggl[2M^2 - M(N-2) \biggr] \biggr\} \Phi_{A B}
\end{displaymath}

\begin{displaymath}
= \sum_{m=0}^{M} \sum_{n=0}^{M} \biggl\{ m(\frac{N}{2} - 1 - m)
+ n(\frac{N}{2} - 1 - n)
\end{displaymath}

\begin{equation}
- (\frac{m-n}{2}) \frac{z_{A} + z_{B}}
{z_{A} - z_{B}} \biggr\} z_{A}^{m} z_{B}^{n} P_{m} P_{n}
\; \; \; ,
\end{equation}

\noindent
and thus that

\begin{displaymath}
\frac{1}{2} \biggl\{ \sum_{j}^{M} z_{j}^2 \frac{\partial^2
\Phi_{mn}} {\partial z_{j}^2} + \sum_{j \neq k}^M \frac{4
z_{j}^2} {z_{j}-z_{k}} \frac{\partial \Phi_{mn}}{\partial z_{j}}
\biggr\}
- \frac{N-3}{2} \sum_{j}^{M} z_{j} \frac{\partial \Phi_{mn}}
{\partial z_{j}}
\end{displaymath}

\begin{displaymath}
= \biggl\{  m(\frac{N}{2} - 1 - m) + n(\frac{N}{2} - 1 - n)
+ \frac{m-n}{2} \biggr\} \Phi_{mn}
\end{displaymath}

\begin{equation}
- \sum_{\ell = 0}^{n} (m-n+ 2\ell ) \; \Phi_{m+\ell, n-\ell}
\; \; \; ,
\label{eig3}
\end{equation}

\noindent
for $m \geq n$, where

\begin{equation}
\Phi_{mn } = \sum_{\alpha = 1}^{N} \sum_{\beta = 1}^{N}
(z_{\alpha}^{*})^m (z_{\beta}^{*})^n \Phi_{\alpha \beta}
= N^2 P_{m} P_{n} \; \; \; .
\end{equation}

\noindent
In obtaining this last expression we have used the identity

\begin{equation}
\frac{x + y}{x - y} (x^{m} y^{n} - x^{n} y^{m})
= 2 \sum_{\ell = 0}^{m-n} x^{m-\ell} y^{n+\ell}
- (x^{m} y^{n} + x^{n} y^{m}) \; \; \; .
\end{equation}

\noindent
The solution of Eq. (\ref{eig2}) is then

\begin{displaymath}
\Phi = \sum_{\ell = 0}^{n} a_{\ell} \; \Phi_{m+\ell , n-\ell}
\end{displaymath}

\begin{equation}
\lambda =   m(\frac{N}{2} - 1 - m) + n(\frac{N}{2} - 1 - n)
- \frac{m-n}{2} \; \; \; ,
\end{equation}

\noindent
where the coefficients $a_{\ell}$ are given by Eq. (\ref{fmn}).
Such a simple solution is possible because the matrix to which Eq.
(\ref{eig3}) corresponds is lower triangular, i.e. takes the form

\begin{equation}
{\rm Matrix} = \left[ \begin{array}{lllll}
    E_{0}  &   0    &   0    &  0     & \ldots \\
    v_{10} & E_{1}  &   0    &  0     &        \\
    v_{20} & v_{21} & E_{2}  &  0     &        \\
    v_{30} & v_{31} & v_{32} & E_{3}  & \ldots \\
    \vdots &        &        & \vdots &        \end{array} \right]
\; \; \; .
\label{hfmn}
\end{equation}

\noindent
The eigenvalues of such a matrix are its diagonal elements, and the
corresponding eigenvectors are generated by recursion.  It should
be noted that the upper bound on the sum in Eq. (\ref{fmn}) is
flexible, as $\Phi_{mn}$ is identically zero unless $0 \leq m , n
\leq M \pmod{N}$.  We have chosen the largest possible value so as
to optimize the smoothness and short-rangeness of $f_{mn}(z)$. $\Box$

\bigskip

\begin{figure}
\epsfbox{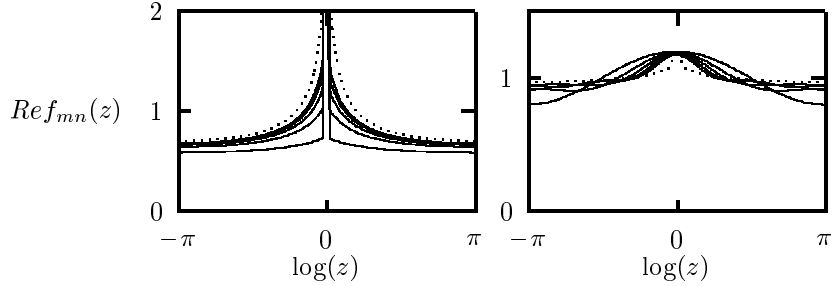}
\caption{Left:  Real part of spinon pair internal wavefunction
         $f_{mn}(z )$ defined by Eq. (50) for the case of $N =
         100$ and $m - n = 2, 4, ... , 10$ (solid) and 50 (dots).
         Right:  The ``adjoint'' function $\bar{f}_{mn}(z)$
         defined in Problem 7.}
\end{figure}

It may be seen in Fig. 5 that $f_{mn}(z_\alpha / z_\beta)$ exhibits
a scattering resonance, an enhancement when $z_\alpha^* z_\beta \simeq
1$, indicating that the spinons attract each other.  This attractive
force may also be inferred from the energy eigenvalue if we rewrite
it as

\begin{equation}
{\cal H}_{HS} | \Psi_{mn} \! > = \biggl\{ - J (\frac{\pi^2}{24}) ( N +
\frac{5}{N} ) + E_{q_{1}} + E_{q_{2}} + V_{q_{1} - q_{2}} \biggr\}
| \Psi_{mn} \! > \; \; \; ,
\label{eig4}
\end{equation}

\noindent
where

\begin{equation}
q_{1} = \frac{\pi}{2} - \frac{2\pi}{N}(m + \frac{1}{2})
\; \; \; \; \; \; \; \; \; \;
q_{2} = \frac{\pi}{2} - \frac{2\pi}{N}(n + \frac{1}{2})
\end{equation}

\noindent
are the spinon momenta, $E_{q}$ is defined as in Eq. (\ref{dispersion}),
and

\begin{equation}
V_{q} = - J \frac{\pi}{N} \mid \! q \! \mid
\; \; \; .
\end{equation}

\noindent
Note that this potential vanishes as $N \rightarrow \infty$, as
expected for particles that interact only when they are close together,
and that the total crystal momentum, as defined by Eq. (\ref{crys2}),
is

\begin{equation}
q = \frac{\pi}{2}(N - 2) + q_1 + q_2 \; \; \pmod{2\pi} \; \; \; ,
\end{equation}

\noindent
a value greater by $\pi$ than that of the ground state when
$q_1 = q_2 = 0$.

      For every triplet state $| \Psi_{mn} \! >$ defined as in Eq.
(\ref{triplet}) there is a corresponding singlet $\Lambda^{z} S^{+}
| \Psi_{mn} \! >$, where $\vec{\Lambda}$ is defined as in Eq.
(\ref{lambda}), with exactly the same energy eigenvalue.  For example,
for the case of $ m = N/2 - 1$ and $n = 0$ we have

\begin{equation}
[ \Lambda^z S^+ \Psi_{mn}] (z_1 , ... , z_{N/2}) = - N^2
\prod_{j<k}^{N/2} (z_j - z_k)^2 \biggl[1 - \prod_{j=1}^{N/2} z_j^2
\biggr] \; \; \; ,
\end{equation}

\noindent
which may be seen to be the alternate ground state defined in Eq.
(\ref{alternate}).

\bigskip

\begin{center}
{\bf Proof}
\end{center}

\noindent
We begin by using the sum rules

\begin{equation}
\prod_j^M z_j \prod_j^{N-M} \eta_j = 1
\; \; \; \; \; \; \; \; \; \; \;
z_j \prod_{k \neq j}^M (z_j - z_k) \prod_k^{N-M} (z_j - \eta_k) = N
\; \; \; ,
\end{equation}

\noindent
to rewrite the wavefunction in terms of the $\downarrow$ spin locations
$\eta_j$, per

\begin{equation}
\Psi_{mn}(z_1 , ... , z_{N/2-1})
= \prod_{j<k}^{N/2-1} (z_j - z_k)^2 \prod_j^{N/2-1} z_j^2
= \prod_{j<k}^{N/2+1} (\eta_j - \eta_k)^2
\; \; \; .
\end{equation}

\noindent
We then have

\begin{displaymath}
[ S^+ \Psi_{mn}] (z_1 , ... , z_{N/2}) = \sum_\alpha
\Psi_{mn} (\eta_1 , ... , \eta_{N/2}, z_\alpha)
\end{displaymath}

\begin{equation}
= N \prod_{j<k}^{N/2}
(\eta_j - \eta_k)^2 \biggl[ 1 + \prod_{j=1}^{N/2} \eta_j^2 \biggr]
= N \prod_{j<k}^{N/2}
(z_j - z_k)^2 \biggl[ 1 + \prod_{j=1}^{N/2} z_j^2 \biggr]
\; \; \; ,
\end{equation}

\noindent
and thus

\begin{displaymath}
[\Lambda^z S^+ \Psi_{mn}] (z_1 , ... , z_{N/2})
\end{displaymath}

\begin{displaymath}
= \frac{1}{2} \sum_{\alpha \neq \beta} ( \frac{z_\alpha + z_\beta}
{z_\alpha - z_\beta}) [S_\alpha^+ S_\beta^- S^+
\Psi_{mn}](z_1 , ... , z_{N/2})
\end{displaymath}

\begin{displaymath}
= \frac{1}{2}
\sum_j \sum_{\beta \neq j} ( \frac{z_j + z_\beta}{z_j - z_\beta})
[S^+ \Psi_{mn}](z_1 , ... , z_{j-1}, z_\beta , z_{j+1} , ... , z_{N/2})
\end{displaymath}

\begin{displaymath}
= N \sum_j^{N/2} z_j \frac{\partial}{\partial z_j}
\biggl\{ \prod_{j<k}^{N/2} (z_j - z_k)^2 \biggl[ 1 +
\prod_{j=1}^{N/2} z_j^2 \biggr] \biggr\}
\end{displaymath}

\begin{equation}
= - N^2 \prod_{j<k}^{N/2} (z_j - z_k)^2 \biggl[ 1 -
\prod_{j=1}^{N/2} z_j^2 \biggr] \; \; \; . \; \Box
\end{equation}

\bigskip

\noindent
The singlet has no simple exact representation in terms of
$| \Psi_{\alpha \beta} \! >$ but is reasonably approximated by

\begin{displaymath}
\Lambda^z S^+ | \Psi_{mn} \! >
\end{displaymath}

\begin{equation}
\simeq \frac{1}{2}
\sum_{\alpha \neq \beta}^{N} \; f_{mn}(z_{\alpha}
/ z_{\beta}) \; (z_{\alpha}^{*})^{m} (z_{\beta}^{*})^{n}
( \frac{z_\alpha + z_\beta}{z_\alpha - z_\beta} ) (S_\alpha^+
- S_\beta^+ ) | \Psi_{\alpha \beta} \! > \; \; \;.
\end{equation}

\noindent
That it is an energy eigenstate follows from the conservation of
$\vec{\Lambda}$ and the non-commutativity of $\vec{\Lambda}$ and
$\vec{S}$ implicit in Eqs. (\ref{vector}).  That it is a singlet
follows from

\begin{equation}
\Lambda^{z} |\Psi_{mn} \! > = \biggl\{ \frac{N - 2}{2}
- m - n \biggr\} |\Psi_{mn} \!> \; \; \; ,
\label{current2}
\end{equation}

\noindent
i.e. that  $|\Psi_{mn} \! >$ is an eigenstate of $\Lambda^z$, an
important result we shall revisit.  This implies that the spin-2
representation contained in the 9 states $\Lambda^\mu S^\nu |
\Psi_{mn} \! > \; \; (\mu , \nu = 1 , 2 , 3$) must be identically
zero and the spin-1 representation must be just $| \Psi_{mn} \! >$
itself.

Spinons are semions, i.e. particles obeying 1/2 fractional statistics.
Since the 2-spinon wavefunction

\begin{equation}
\Psi_{AB} = \prod_j (z_j - z_A )( z_j - z_B ) \prod_{j < k}
(z_j - z_k)^2 \; \; \; ,
\end{equation}

\noindent
where $z_A$ and $z_B$ are not necessarily lattice sites, has the
property

\begin{equation}
\Psi_{AB}^* (z_1 , ... , z_{N/2-1}) = (z_A z_B)^{1 - N/2}
\Psi_{AB} (z_1 , ... , z_{N/2-1}) \; \; \; ,
\end{equation}

\noindent
the Berry phase vector potential for adiabatic motion of spinon A
in the presence of B is

\begin{displaymath}
\frac{1}{2} \biggl[
< \! \psi_{AB} | z_A \frac{\partial} {\partial z_A} \psi_{AB} \! > +
< \! z_A \frac{\partial} {\partial z_A} \psi_{AB} | \psi_{AB} \! >
\biggr] / < \! \psi_{AB} | \psi_{AB} \! >
\end{displaymath}

\begin{equation}
= \frac{1}{2} (1 - \frac{N}{2}) \; \; \; .
\end{equation}

\noindent
The phase to ``exchange'' the spinons by moving $A$ all the way
around the loop is thus

\begin{equation}
\Delta \phi = \oint \frac{1}{2} (1 - \frac{N}{2}) \frac{dz_A}{z_A}
= \pm \frac{\pi}{2} i \; \; \; \pmod{2\pi} \; \; \; .
\end{equation}

\noindent
This number is $0$ or $\pi$ for bosons or fermions.
Fractional statistics is actually the long-range force between
the spinons manifested in the resonant enhancements of Fig. 5 and
the potential $V_{q_1 - q_2}$ in Eq. (\ref{eig4}), and has nothing
to do with the symmetry or antisymmetry of $|\Psi_{\alpha \beta}
\! >$ under interchange of $\alpha$ and $\beta$.  It does have to do
with state-counting. The number of states available to
$\ell$ $\downarrow$ spinons, determined by counting the number
of distinct symmetric polynomials of the form

\begin{equation}
\Phi_{z_{A_1} , ... , z_{A_\ell}} (z_1 , ... , z_{(N-\ell)/2})
= \prod_j^{(N - \ell) / 2} (z_j - z_{A_1}) \times ... \times
(z_j - z_{A_\ell}) \; \; \; ,
\end{equation}

\noindent
is

\begin{equation}
{\cal N}_\ell^{\rm semi} =
( \! \! \begin{array}{c} N/2 + \ell/2 \\ \ell \end{array} \! \! )
\; \; \; .
\end{equation}

\noindent
This is just halfway between the numbers

\begin{equation}
{\cal N}_\ell^{\rm fermi} =
( \! \! \begin{array}{c} N/2 \\ \ell \end{array} \! \! )
\; \; \; \; \; \; \; \; \; \; \; \; \; \; \; \;
{\cal N}_\ell^{\rm bose} =
( \! \! \begin{array}{c} N/2 + \ell \\ \ell \end{array} \! \! )
\; \; \; ,
\end{equation}

\noindent
likewise calculated assuming that the number of states available for
one particle is $N/2$.

\section{Annihilation Operators}

The operators

\begin{equation}
\vec{\Omega}_\alpha = \frac{1}{2} \sum_{\beta \neq \alpha} \biggl(
\frac{z_\alpha + z_\beta} {z_\alpha - z_\beta} \biggr)
[ i (\vec{S}_\alpha \times \vec{S}_\beta) + \vec{S}_\beta ]
\end{equation}

\noindent
annihilate the Haldane-Shastry ground state, i.e. satisfy

\begin{equation}
\vec{\Omega}_\alpha | \Psi \! > = 0 \; \; \; ,
\end{equation}

\noindent
for all $\alpha$.

\bigskip

\begin{center}
{\bf Proof}
\end{center}

\noindent
We have as before that $[S_\alpha^+ S_\beta^- \Psi ] (z_1 , ... ,
z_{N/2})$ is zero unless one of the arguments $z_1 , ... , z_{N/2}$
equals $z_\alpha$, but in this case we do not sum over $z_\alpha$.
Instead we have

\begin{displaymath}
\sum_{\beta \neq \alpha}^N \frac{z_\alpha}{z_\alpha - z_\beta}
[ S_\alpha^+ S_\beta^- \Psi ] (z_\alpha , z_2 , ... , z_{N/2})
= \sum_{\beta \neq \alpha}^N \frac{z_\alpha}{z_\alpha - z_\beta}
\Psi (z_\beta , z_2 , ... , z_{N/2})
\end{displaymath}

\begin{displaymath}
= \sum_{\ell = 0}^{N-2} \biggl\{\frac{1}{\ell!}
\sum_{\beta \neq \alpha}^N \frac{z_\alpha z_\beta
(z_\beta - z_\alpha )^\ell} {z_\alpha - z_\beta} \biggr\}
\frac{\partial^\ell}{\partial z_\alpha^\ell} \biggl\{
\frac{\Psi (z_\alpha , z_2 , ... , z_{N/2} )}{z_\alpha} \biggr\}
\end{displaymath}

\begin{equation}
= \biggl\{ - \frac{N-1}{2} + 2 \sum_{j \neq \alpha}^{N/2}
\frac{z_\alpha} {z_\alpha - z_j} \biggr\} \Psi (z_\alpha ,
z_2 , ... , z_{N/2}) \; \; \; .
\end{equation}

\noindent
However $(1/2 + S_\alpha^z ) | \Psi \! >$ is also identically zero
unless one of the arguments $z_1 , ... , z_{N/2}$ equals $z_\alpha$.
We thus have

\begin{displaymath}
\sum_{\beta \neq \alpha}^N [ \frac{z_\alpha}{z_\alpha - z_\beta}
( \frac{1}{2} + S_\alpha^z ) (\frac{1}{2} + S_\beta^z ) \Psi ]
(z_\alpha , z_2 , ... , z_{N/2})
\end{displaymath}

\begin{equation}
= \sum_{j \neq \alpha}
\frac{z_\alpha}{z_\alpha - z_j} \Psi (z_\alpha , z_2 , ... ,
z_{N/2}) \; \; \; .
\end{equation}

\noindent
Subtracting these from each other we find that

\begin{displaymath}
\biggl\{ \sum_{\beta \neq \alpha}^N \frac{z_\alpha}
{z_\alpha - z_\beta} \biggl[ S_\alpha^+ S_\beta^- - 2 (\frac{1}{2}
+ S_\alpha^z ) (\frac{1}{2} + S_\beta^z ) \biggr]
\end{displaymath}

\begin{equation}
+ \frac{N-1}{2}
(\frac{1}{2} + S_\alpha^z ) \biggr\} | \Psi \! > = 0
\end{equation}

\noindent
for all $\alpha$.  However since $| \Psi \! >$ is a spin singlet
the irreducible representations of the rotation group present
in this operator must destroy $| \Psi \! >$ separately.  The scalar
component is identically zero.  The vector component is

\begin{equation}
\sum_{\beta \neq \alpha}^N \frac{z_\alpha}{z_\alpha - z_\beta}
[ i (\vec{S}_\alpha \times \vec{S}_\beta ) + \vec{S}_\beta ]
| \Psi \! > = 0 \; \; \; .
\end{equation}

\noindent
The rank-2 tensor component is the product of the two and
therefore also zero.  Since $| \Psi \! >$ is also its own
time-reverse it must be destroyed by the time-reverse of
the vector operator, i.e.

\begin{equation}
\sum_{\beta \neq \alpha} \frac{z_\alpha^*}
{z_\alpha^* - z_\beta^*} [ i (\vec{S}_\alpha \times \vec{S}_\beta )
+ \vec{S}_\beta ] = - \sum_{\beta \neq \alpha} \frac{z_\beta}
{z_\alpha - z_\beta} [ i (\vec{S}_\alpha \times \vec{S}_\beta )
+ \vec{S}_\beta ] \; \; \; .
\end{equation}

\noindent
The difference of these is the trivial operator $\vec{S}_\alpha \times
\vec{S}$, and their sum is $2 \vec{\Omega}_\alpha$. $\Box$

\bigskip

\noindent
These operators satisfy

\begin{equation}
\vec{S}_\alpha \times \vec{\Omega}_\alpha = - \frac{i}{2}
\vec{\Omega}_\alpha
\; \; \; \; \; \; \; \; \; \; \;
\vec{S}_\alpha \cdot \vec{\Omega}_\alpha = 0
\end{equation}

\noindent
and are related to $\vec{\Lambda}$ by

\begin{equation}
\sum_\alpha \vec{\Omega}_\alpha = \vec{\Lambda}
\; \; \; .
\end{equation}

\noindent
They are not symmetries of the Hamiltonian but supercharges, for we
have

\begin{displaymath}
\sum_\alpha \vec{\Omega}_\alpha^\dagger \cdot \vec{\Omega}_\alpha
\end{displaymath}

\begin{equation}
= \frac{3}{2} \biggl[
 3 \sum_{\alpha \neq \beta} \frac{\vec{S}_\alpha \cdot
 \vec{S}_\beta}{|z_\alpha - z_\beta|^2} + \frac{N(N^2 + 5)}{16}
- \frac{(N+1)}{4} S^2 \biggr]
\; \; \; ,
\label{superch}
\end{equation}

\noindent
exactly.

\bigskip

\begin{center}
{\bf Proof}
\end{center}

\noindent
Since

\begin{equation}
\vec{\Omega}_\alpha = \sum_{\beta \neq \alpha} \frac{z_\alpha}{z_\alpha
- z_\beta} [ i (\vec{S}_\alpha \times \vec{S}_\beta ) + \vec{S}_\beta ]
- \frac{1}{2} [ i (\vec{S}_\alpha \times \vec{S}) + \vec{S} ]
\; \; \; ,
\end{equation}

\noindent
per the previous discussion, we have

\begin{displaymath}
\sum_\alpha \sum_{\beta \neq \alpha} \sum_{\gamma \neq \alpha} \frac{
[i (\vec{S}_\alpha \times \vec{S}_\gamma) + \vec{S}_\gamma ]^\dagger
\cdot [i (\vec{S}_\alpha \times \vec{S}_\beta) + \vec{S}_\beta ]}
{(z_\alpha^* - z_\gamma^*)(z_\alpha - z_\beta )}
\end{displaymath}

\begin{equation}
= \sum_\alpha \vec{\Omega}_\alpha^\dagger \cdot \vec{\Omega}_\alpha
+ \frac{3}{2} \vec{S} \cdot \vec{\Lambda} + \frac{3}{8} (N - 1) S^2
\; \; \; .
\end{equation}

\noindent
In evaluating this expression we have used

\begin{equation}
[i (\vec{S}_\alpha \times \vec{S}_\gamma ) + \vec{S}_\gamma ]^\dagger
= [i (\vec{S}_\gamma \times \vec{S}_\alpha ) + \vec{S}_\gamma ]
\; \; \; ,
\end{equation}

\begin{equation}
\sum_\alpha [i (\vec{S} \times \vec{S}_\alpha) + \vec{S} ] \cdot
\vec{\Omega}_\alpha = \sum_\alpha [i \vec{S} \cdot (\vec{S}_\alpha
\times \vec{\Omega}_\alpha ) + \vec{S} \cdot \vec{\Omega}_\alpha ]
= \frac{3}{2} \vec{S} \cdot \vec{\Lambda} \; \; \; ,
\end{equation}

\noindent
and

\begin{displaymath}
\sum_\alpha [ i ( \vec{S} \times \vec{S}_\alpha ) + \vec{S} ] \cdot
[ i (\vec{S}_\alpha \times \vec{S} ) + \vec{S} ]
\end{displaymath}

\begin{equation}
= \frac{3}{2} \sum_\alpha [ S^2 - i \vec{S} \cdot (\vec{S}_\alpha
\times \vec{S} ) ] = \frac{3}{2} [ N - 1 ] S^2 \; \; \; ,
\end{equation}

\noindent
which follows from

\begin{displaymath}
[ i(\vec{S}_\gamma \times \vec{S}_\alpha) + \vec{S}_\gamma ]
\cdot [ i (\vec{S}_\alpha \times \vec{S}_\beta) + \vec{S}_\beta ]
\end{displaymath}

\begin{displaymath}
=- (\vec{S}_\gamma \times \vec{S}_\alpha) \cdot (\vec{S}_\alpha \times
\vec{S}_\beta) + i \vec{S}_\gamma \cdot (\vec{S}_\alpha \times
\vec{S}_\beta) + i (\vec{S}_\gamma \times \vec{S}_\alpha) \cdot
\vec{S}_\beta + \vec{S}_\gamma \cdot \vec{S}_\beta
\end{displaymath}

\begin{displaymath}
= (1 + \frac{3}{4} ) ( \vec{S}_\gamma \cdot \vec{S}_\beta ) - ( \vec{S}_\gamma
\cdot \vec{S}_\alpha ) ( \vec{S}_\alpha \cdot \vec{S}_\beta ) + 2i \vec{S}_\gamma \cdot
( \vec{S}_\alpha \times \vec{S}_\beta )
\end{displaymath}

\begin{equation}
= \frac{3}{2} [ \vec{S}_\gamma \cdot \vec{S}_\beta + i \vec{S}_\gamma \cdot
( \vec{S}_\alpha \times \vec{S}_\beta)]
\; \; \; .
\end{equation}

\noindent
The 2-spin sum is

\begin{displaymath}
\sum_{\beta \neq \gamma \neq \alpha} \frac{
\vec{S}_\beta \cdot \vec{S}_\gamma}{ ( z^*_\alpha - z^*_\gamma )( z_\alpha -
z_\beta ) } = -\sum_{\alpha \neq \beta \neq \gamma} \frac{z_\alpha
z_\gamma} {(z_\alpha - z_\gamma)(z_\alpha - z_\beta)} \vec{S}_\gamma
\cdot \vec{S}_\beta
\end{displaymath}

\begin{displaymath}
= - \sum_{\beta \neq \gamma} \frac{z_\gamma}{ z_\beta - z_\gamma}
\vec{S}_\gamma \cdot \vec{S}_\beta \sum_{\alpha \neq \beta, \gamma} \biggl[
\frac{z_\alpha}{z_\alpha - z_\beta} - \frac{z_\alpha}{z_\alpha -
z_\gamma} \biggr]
\end{displaymath}

\begin{displaymath}
= - \sum_{\beta \neq \gamma} \frac{z_\gamma}{z_\beta - z_\gamma}
\vec{S}_\gamma \cdot \vec{S}_\beta \biggl[ \frac{z_\beta}{z_\beta - z_\gamma}
- \frac{z_\gamma}{z_\gamma - z_\beta} \biggr]
\end{displaymath}

\begin{displaymath}
= -\sum_{\beta \neq \gamma} \frac{z_\gamma (z_\gamma + z_\beta)}
{(z_\beta - z_\gamma)^2} \vec{S}_\gamma \cdot \vec{S}_\beta
= -\frac{1}{2} \sum_{\beta \neq \gamma} \biggl( \frac{z_\beta +
z_\gamma}{z_\beta - z_\gamma} \biggr)^2 \vec{S}_\beta \cdot \vec{S}_\gamma
\end{displaymath}

\begin{equation}
= -\frac{1}{2} S^2 + \frac{3}{8} N + 2\sum_{\alpha \neq \beta}
\frac{\vec{S}_\alpha \cdot \vec{S}_\beta}{|z_\alpha - z_\beta|^2}
\; \; \; .
\end{equation}

\noindent
The 3-spin sum is

\begin{displaymath}
i\sum_{\alpha \neq \beta \neq \gamma} \frac{\vec{S}_\gamma \cdot
(\vec{S}_\alpha \times \vec{S}_\beta)}{(z^*_\alpha - z^*_\gamma) ( z_\alpha -
z_\beta)}
\end{displaymath}

\begin{displaymath}
= i \sum_{\alpha \neq \beta \neq \gamma} \frac{z_\alpha z_\gamma}
{ ( z_\alpha - z_\gamma ) ( z_\alpha - z_\beta ) }
\vec{S}_\alpha \cdot (\vec{S}_\gamma \times
\vec{S}_\beta)
\end{displaymath}

\begin{displaymath}
= \frac{i}{2} \sum_{\alpha \neq \beta \neq \gamma} \frac{z_\alpha
( z_\gamma - z_\beta)}{(z_\alpha - z_\beta) ( z_\alpha - z_\gamma)}
\vec{S}_\alpha \cdot (\vec{S}_\gamma \times \vec{S}_\beta)
\end{displaymath}

\begin{displaymath}
= \frac{i}{2} \sum_{\alpha \neq \beta \neq \gamma} \biggl[
\frac{z_\alpha}{z_\alpha - z_\gamma} - \frac{z_\alpha}{z_\alpha -
z_\beta} \biggr] \vec{S}_\alpha \cdot (\vec{S}_\gamma \times \vec{S}_\beta)
\end{displaymath}

\begin{equation}
= \frac{i}{2} \sum_{\alpha \neq \beta \neq \gamma} \biggl(
\frac{z_\alpha + z_\gamma}{z_\alpha - z_\gamma} \biggr)
(\vec{S}_\alpha \times \vec{S}_\gamma) \cdot \vec{S}_\beta
= \vec{\Lambda} \cdot \vec{S}
\; \; \; .
\end{equation}

\noindent
Note that in the last step we have used the identity

\begin{equation}
(\vec{S}_\alpha \times \vec{S}_\gamma) \cdot (\vec{S}_\alpha +
\vec{S}_\gamma) = 0
\; \; \; .
\end{equation}

\noindent
Putting these results together, we find that

\begin{displaymath}
\sum_{\alpha} \sum_{\beta \neq \alpha} \sum_{\gamma \neq \alpha}
\frac{1} {(z^*_\alpha - z^*_\gamma)(z_\alpha - z_\beta)}
[ -i(\vec{S}_\alpha \times \vec{S}_\gamma) + \vec{S}_\gamma] \cdot
[i (\vec{S}_\alpha \times \vec{S}_\beta) + \vec{S}_\beta ]
\end{displaymath}

\begin{displaymath}
= \frac{3}{2} \sum_\alpha \sum_{\beta \neq \alpha} \sum_{\gamma \neq
\alpha} \frac{1}{(z^*_\alpha - z^*_\gamma)(z_\alpha - z_\beta)}
[ \vec{S}_\gamma \cdot \vec{S}_\beta + i \vec{S}_\gamma \cdot
(\vec{S}_\alpha \times \vec{S}_\beta)]
\end{displaymath}

\begin{displaymath}
= \frac{3}{2} \biggl\{ \sum_{\beta \neq \alpha } \frac{1}{|z_\alpha -
z_\beta|^2} \biggl[ \frac{3}{4} + 3 \; \vec{S}_\alpha \cdot \vec{S}_\beta
\biggr] - \frac{S^2}{2} + \frac{3}{8} N + \vec{\Lambda} \cdot \vec{S}
\biggr\}
\end{displaymath}

\begin{equation}
= \frac{3}{2} \biggl[ 3 \sum_{\alpha \neq \beta} \frac{\vec{S}_\alpha
\cdot \vec{S}_\beta}{|z_\alpha - z_\beta|^2} + \frac{N(N^2 + 5)}{16}
- \frac{S^2}{2} + \vec{S} \cdot \vec{\Lambda} \biggr]
\; \; \; . \; \; \Box
\end{equation}

\bigskip

\noindent
Since $< \! \Phi | \vec{\Omega}_\alpha^\dagger \cdot
\vec{\Omega}_\alpha | \Phi \! >$ is non-negative for any wavefunction
$| \Phi \! >$, this provides an explicit demonstration that
$| \Psi \! >$ is the true ground state.  The annihilation operators
and their equivalence to ${\cal H}_{HS}$ when squared and summed were
originally discovered by Shastry \cite{shastry1}.  They are modeled
after a similar set of operators discovered for the 2-dimensional
chiral spin liquid, although there was a minus-sign error in the
original paper which caused the operators to be
mistakenly reported as scalars \cite{rbl}.  They are lattice
versions of the Knizhnik-Zamolodchikov operators known from studies
of the Cologero-Sutherland model, the 1-dimensional Bose gas with
inverse-square repulsions \cite{zam,cologero}.

\section{Spin Current}

The operator $\vec{\Lambda}$ is a scaled spin current.  Its action
on the propagating spinon of Eq. (\ref{spinon}), for example, is

\begin{equation}
\Lambda^z | \Psi_m \! > = \biggl\{ \frac{N-1}{4} - m \biggr\}
| \Psi_m \! > \; \; \; ,
\end{equation}

\noindent
which is proportional to the spinon velocity

\begin{equation}
\frac{dE_q}{dq} = \frac{2 \pi J}{N} \biggl\{ \frac{N-1}{4} - m
\biggr\} \; \; \; .
\end{equation}

\noindent
Its action on the 2-spinon state given by Eq. (\ref{current2}) is
similarly the sum of the two spinon velocities.

\bigskip

\begin{center}
{\bf Proof}
\end{center}

\noindent
We have, with $M = (N-1)/2$,

\begin{displaymath}
[ \Lambda^z \Psi_A ] (z_1 , ... , z_M)
= \frac{1}{2} \sum_j^M \sum_{\beta \neq j} ( \frac{z_j + z_\beta}
{z_j - z_\beta} ) \Psi_A ( z_1 , ... , z_{j-1} , z_\beta , z_{j+1},
... , z_M )
\end{displaymath}

\begin{displaymath}
= \frac{1}{2} \sum_j^M \sum_\ell \frac{1}{\ell !} \biggl[
\sum_{\beta \neq j} ( \frac{z_j + z_\beta}{z_j - z_\beta} )
(z_\beta - z_j)^\ell z_\beta \biggr] \frac{\partial^\ell}
{\partial z_j^\ell} \biggl\{ \Psi_A (z_1 , ... , z_M) / z_j \biggr\}
\end{displaymath}

\begin{equation}
= \biggl\{ \frac{N-1}{4} - z_A \frac{\partial}{\partial z_A}
\biggr\} \Psi_A (z_1 , ... , z_M ) \; \; \; ,
\end{equation}

\noindent
and thus

\begin{equation}
\Lambda^z | \Psi_m \! > = \sum_{z_A} (z_A^*)^m \Lambda^z | \Psi_A \! >
= \biggl\{ \frac{N-1}{4} - m \biggr\} | \Psi_m \! > \; \; \; .
\; \; \Box
\end{equation}

\bigskip

\noindent
A more traditional description of this current may be constructed by
interpolating the spin operators into the interstices by means of the
formula

\begin{equation}
\vec{\sigma}(z) = \biggl[ \frac{z^{N/2} - z^{-N/2}}{2N}\biggr]
\sum_\beta \biggl( \frac{z + z_\beta}{z - z_\beta}
\biggr) \vec{S}_\beta \; \; \; .
\label{sigma}
\end{equation}

\noindent
The Hamiltonian is then

\begin{equation}
\frac{1}{2\pi i} \oint
[ z \frac{d \vec{\sigma}}{d z} ] \cdot [ z \frac{d \vec{\sigma}} {d z} ]
\frac{dz}{z} = - \frac{2}{N} \sum_{\alpha \neq \beta}^N
\frac{\vec{S}_\alpha \cdot \vec{S}_\beta}{| z_\alpha - z_\beta |^2}
+ \frac{3}{8}(N - 1) + \frac{S^2}{8} \; \; \; ,
\end{equation}

\noindent
where the integral is performed over the unit circle.  We also have
spin density and spin current density operators

\begin{displaymath}
\vec{\rho}(z) = - i \vec{\sigma}(z) \times \vec{\sigma}(z)
\end{displaymath}

\begin{equation}
\vec{j} (z) =
\frac{1}{2i} \biggl\{ \vec{\sigma} \times [ z \frac{d \vec{\sigma}}{d z}]
- [ z \frac{d \vec{\sigma}}{d z} ] \times \vec{\sigma} \biggr\}
\; \; \; ,
\end{equation}

\noindent
which satisfy the continuity equation

\begin{equation}
\lim_{z \rightarrow z_\alpha} \; \biggl\{  z \frac{d \vec{j}}
{d z} + [ \sum_{\alpha \neq \beta} \frac{\vec{S}_\alpha \cdot
\vec{S}_\beta}{| z_\alpha - z_\beta |^2 }
, \vec{\rho} ] \biggr\} = 0
\; \; \; .
\end{equation}

\noindent
The zero-momentum component of this conserved current density is

\begin{equation}
\frac{1}{2\pi i} \oint \vec{j} \; \frac{dz}{z} = \vec{\Lambda}
\; \; \; .
\end{equation}

\section{Supersymmetry}

We shall now consider the generalization of the Haldane-Shastry
Hamiltonian

\begin{displaymath}
{\cal H}_{KY} = J \left( \frac{2\pi}{N} \right)^2
\sum_{\alpha < \beta}^N \frac{1}{ | z_\alpha - z_\beta |^2 } \;
P \biggl\{ \vec{S}_\alpha \cdot \vec{S}_\beta
\end{displaymath}

\begin{equation}
- \frac{1}{4} \sum_s
( c_{\alpha s}^\dagger c_{\beta s} + c_{\beta s}^\dagger c_{\alpha s} )
+ \frac{1}{2} (n_\alpha + n_\beta) - \frac{1}{4} n_\alpha n_\beta
- \frac{3}{4} \biggr\} P
\end{equation}

\noindent
first studied by Kuramoto and Yokoyama \cite{kuramoto}, where

\begin{equation}
P = \prod_\alpha \; ( 1 -
c_{\alpha \uparrow}^\dagger c_{\alpha \downarrow}^\dagger
c_{\alpha \downarrow} c_{\alpha \uparrow} )
\; \; \; ,
\label{gutz}
\end{equation}

\noindent
is the Gutzwiller projector, and site occupation and spin operators
are

\begin{displaymath}
n_\alpha =  c_{\alpha \uparrow}^\dagger c_{\alpha \uparrow}
+ c_{\alpha \downarrow}^\dagger c_{\alpha \downarrow}
\; \; \; \; \; \; \; \; \; \; \;
S_\alpha^x = \frac{1}{2} (
c_{\alpha \uparrow}^\dagger c_{\alpha \downarrow} +
c_{\alpha \downarrow}^\dagger c_{\alpha \uparrow} )
\end{displaymath}

\begin{equation}
S_\alpha^y = \frac{1}{2i} (
c_{\alpha \uparrow}^\dagger c_{\alpha \downarrow} -
c_{\alpha \downarrow}^\dagger c_{\alpha \uparrow} )
\; \; \; \; \; \; \; \;
S_\alpha^z = \frac{1}{2} (
c_{\alpha \uparrow}^\dagger c_{\alpha \uparrow} -
c_{\alpha \downarrow}^\dagger c_{\alpha \downarrow} )
\; \; \; .
\end{equation}

\noindent
Thus each site can be $\uparrow$, $\downarrow$, or  unoccupied - but
not doubly occupied - and the hole can tunnel to nearby sites by means
of the same inverse-square matrix element characterizing the spin
exchange.  This equivalence of the energy scales for magnetism and
charge transport causes the Hamiltonian to be supersymmetric in the
sense of Eq. (\ref{superch}) and also in the more traditional one of
commuting with electron or hole injection per

\begin{equation}
\sum_\alpha \; [ {\cal H}_{KY} , P  c_{\alpha s} P ] = 0
\; \; \; .
\end{equation}

\bigskip

\begin{center}
{\bf Proof}
\end{center}

\noindent
It suffices to show that

\begin{equation}
[ {\cal H}_{\alpha \beta} , (c_{\alpha \uparrow} + c_{\beta \uparrow}) ]
\; | \psi \! > = 0 \; \; \; ,
\end{equation}

\noindent
where

\begin{displaymath}
{\cal H}_{\alpha \beta}
= P  \biggl\{ \vec{S}_\alpha \cdot \vec{S}_\beta
- \frac{1}{4} \sum_s
( c_{\alpha s}^\dagger c_{\beta s} + c_{\beta s}^\dagger c_{\alpha s} )
\end{displaymath}

\begin{equation}
- \frac{1}{4} n_\alpha n_\beta + \frac{1}{2} (n_\alpha + n_\beta - 1)
\biggr\} P
\end{equation}

\noindent
for the 9 configurations $| \psi \! >$  on sites $\alpha$ and $\beta$
allowed by the projector $P$.  Denoting the state with no electron
on either site by $| 0 \! >$ we have

\begin{itemize}

\item {\bf Case 1:}

      \begin{equation}
      | \psi_1 \! > = c_{\alpha \uparrow}^\dagger
      c_{\beta \uparrow}^\dagger | 0 \! >
      \end{equation}

      \begin{displaymath}
      (c_{\alpha \uparrow} + c_{\beta \uparrow} ){\cal H}_{\alpha \beta}
      | \psi_1 \! > = \frac{1}{2}
      (c_{\alpha \uparrow} + c_{\beta \uparrow} ) | \psi_1 \! >
      = \frac{1}{2}(c_{\beta \uparrow}^\dagger - c_{\alpha \uparrow}^\dagger )
      | 0 \! >
      \end{displaymath}

      \begin{displaymath}
      {\cal H}_{\alpha \beta} (c_{\alpha \uparrow} + c_{\beta \uparrow} )
      | \psi_1 \! > = {\cal H}_{\alpha \beta}
      (c_{\beta \uparrow}^\dagger - c_{\alpha \uparrow}^\dagger ) | 0 \! >
      = \frac{1}{2}(c_{\beta \uparrow}^\dagger -
      c_{\alpha \uparrow}^\dagger ) | 0 \! >
      \end{displaymath}

\item {\bf Case 2:}

      \begin{equation}
      | \psi_2 \! >
      = ( c_{\alpha \uparrow}^\dagger c_{\beta \downarrow}^\dagger
      - c_{\alpha \downarrow}^\dagger c_{\beta \uparrow}^\dagger )
      | 0 \! >
      \end{equation}

      \begin{displaymath}
      (c_{\alpha \uparrow} + c_{\beta \uparrow} ){\cal H}_{\alpha \beta}
      | \psi_2 \! > = - \frac{1}{2}
      (c_{\alpha \uparrow} + c_{\beta \uparrow} ) | \psi_2 \! >
      =  - \frac{1}{2}(c_{\beta \downarrow}^\dagger +
      c_{\alpha \downarrow}^\dagger )
      | 0 \! >
      \end{displaymath}

      \begin{displaymath}
      {\cal H}_{\alpha \beta} (c_{\alpha \uparrow} + c_{\beta \uparrow} )
      | \psi_2 \! > = {\cal H}_{\alpha \beta}
      (c_{\beta \downarrow}^\dagger + c_{\alpha \downarrow}^\dagger )
      | 0 \! > = - \frac{1}{2}(c_{\beta \downarrow}^\dagger +
      c_{\alpha \downarrow}^\dagger ) | 0 \! >
      \end{displaymath}

\item {\bf Case 3:}

      \begin{equation}
      | \psi_3 \! >
      = ( c_{\alpha \uparrow}^\dagger c_{\beta \downarrow}^\dagger
      + c_{\alpha \downarrow}^\dagger c_{\beta \uparrow}^\dagger )
      | 0 \! >
      \end{equation}

      \begin{displaymath}
      (c_{\alpha \uparrow} + c_{\beta \uparrow} ){\cal H}_{\alpha \beta}
      | \psi_3 \! > = \frac{1}{2}
      (c_{\alpha \uparrow} + c_{\beta \uparrow} ) | \psi_3 \! >
      =  \frac{1}{2}(c_{\beta \downarrow}^\dagger -
      c_{\alpha \downarrow}^\dagger )
      | 0 \! >
      \end{displaymath}

      \begin{displaymath}
      {\cal H}_{\alpha \beta} (c_{\alpha \uparrow} + c_{\beta \uparrow} )
      | \psi_3 \! > = {\cal H}_{\alpha \beta}
      (c_{\beta \downarrow}^\dagger - c_{\alpha \downarrow}^\dagger )
      | 0 \! > = \frac{1}{2}(c_{\beta \downarrow}^\dagger -
      c_{\alpha \downarrow}^\dagger ) | 0 \! >
      \end{displaymath}

\item {\bf Case 4:}

      \begin{equation}
      | \psi_4 \! > = c_{\alpha \uparrow}^\dagger
      c_{\beta \uparrow}^\dagger | 0 \! >
      \end{equation}

      \begin{displaymath}
      (c_{\alpha \uparrow} + c_{\beta \uparrow} ){\cal H}_{\alpha \beta}
      | \psi_4 \! > = \frac{1}{2}
      (c_{\alpha \uparrow} + c_{\beta \uparrow} ) | \psi_4 \! > = 0
      \end{displaymath}

      \begin{displaymath}
      {\cal H}_{\alpha \beta} (c_{\alpha \uparrow} + c_{\beta \uparrow} )
      | \psi_4 \! > = 0
      \end{displaymath}

\item {\bf Cases 5 and 6:}

      \begin{equation}
      | \psi_5 \! > = c_{\alpha \uparrow} | 0 \! >
      \end{equation}

      \begin{displaymath}
      (c_{\alpha \uparrow} + c_{\beta \uparrow} ){\cal H}_{\alpha \beta}
      | \psi_5 \! > = - \frac{1}{2}
      (c_{\alpha \uparrow} + c_{\beta \uparrow} )
      c_{\beta \uparrow}^\dagger | 0 \! > = - \frac{1}{2} | 0 \! >
      \end{displaymath}

      \begin{displaymath}
      {\cal H}_{\alpha \beta} (c_{\alpha \uparrow} + c_{\beta \uparrow} )
      | \psi_5 \! > = {\cal H}_{\alpha \beta} | 0 \! >
      = - \frac{1}{2} | 0 \! >
      \end{displaymath}

\item {\bf Cases 7 and 8:}

      \begin{equation}
      | \psi_7 \! > = c_{\alpha \downarrow} | 0 \! >
      \end{equation}

      \begin{displaymath}
      (c_{\alpha \uparrow} + c_{\beta \uparrow} ){\cal H}_{\alpha \beta}
      | \psi_7 \! > = - \frac{1}{2}
      (c_{\alpha \uparrow} + c_{\beta \uparrow} )
      c_{\beta \downarrow}^\dagger | 0 \! > = 0
      \end{displaymath}

      \begin{displaymath}
      {\cal H}_{\alpha \beta} (c_{\alpha \uparrow} + c_{\beta \uparrow} )
      | \psi_7 \! > = 0
      \end{displaymath}

\item {\bf Case 9:}

      \begin{equation}
      | \psi_9 \! > = | 0 \! >
      \end{equation}

      \begin{displaymath}
      (c_{\alpha \uparrow} + c_{\beta \uparrow} ){\cal H}_{\alpha \beta}
      | \psi_9 \! > = - \frac{1}{2}
      (c_{\alpha \uparrow} + c_{\beta \uparrow} ) | 0 \! > = 0
      \end{displaymath}

      \begin{displaymath}
      {\cal H}_{\alpha \beta} (c_{\alpha \uparrow} + c_{\beta \uparrow} )
      | \psi_9 \! > = 0 \; \; \; . \; \; \; \Box
      \end{displaymath}

\end{itemize}

\bigskip

\begin{figure}
\epsfbox{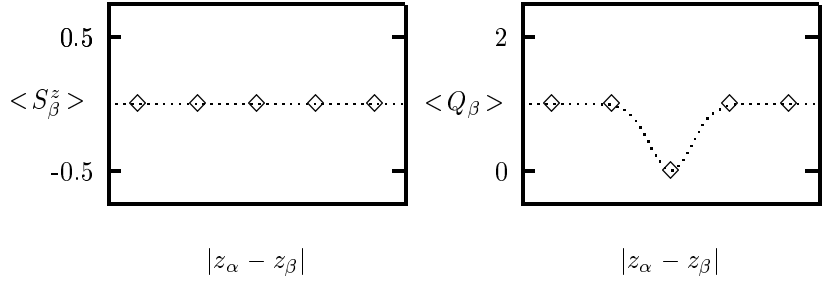}
\caption{Spin and charge profiles of the localized holon
        $c_{a \downarrow} | \Psi_\alpha \! >$.  The dotted lines
        are a guide to the eye.}
\end{figure}

\section{Holons}

Holons are charged, spin-0 elementary excitations of the
Kuramoto-Yokoyama Hamiltonian made by removing an electron from the
center of a spinon. The localized holon $c_{a \downarrow} |
\Psi_\alpha \! >$, where $| \Psi_\alpha \! >$ is defined as in Eq.
(\ref{spinon1}), is shown in Fig. 6.  It is a natural complement to
the spinon from which it was made, for the two states differ only in
the disposition of the central site, which does not fluctuate and is
not involved in any way in the quantum number fractionalization.  This
state is an exact spin singlet by virtue of Eq. (\ref{singlet2}).  Let
$N$ be odd and let $M = (N-1)/2$ as in Eq. (\ref{spinon}).  Then the
propagating holon wavefunction

\begin{equation}
\Psi_m^{holon} (z_1 , ... , z_M | h ) = (h^*)^m \prod_j^M (h - z_j )
\prod_{j < k}^M (z_j - z_k )^2 \prod_j^M z_j
\; \; \; ,
\label{holon}
\end{equation}

\noindent
where $z_1 , ... , z_M$ denote the positions of the $\uparrow$
sites and $h$ denotes the position of the empty site, all others
being $\downarrow$, satisfies

\begin{displaymath}
{\cal H}_{KY} | \Psi_m^{holon} \! >
\end{displaymath}

\begin{equation}
= \biggl\{ -J ( \frac{\pi^2}{24})
(N - \frac{1}{N}) + \frac{J}{2} ( \frac{2\pi}{N} )^2 m (
\frac{N+1}{2} + m ) \biggr\} | \Psi_m^{holon} \! >
\; \; \; ,
\end{equation}

\noindent
provided that $-(N+1)/2 \leq m \leq 0$.

\bigskip

\begin{center}
{\bf Proof}
\end{center}

\noindent
To simplify the notation let us negate the power of $h$ in Eq.
(\ref{holon}) so that the holon wavefunction becomes

\begin{equation}
\Psi_n (z_1 , ... , z_M | h ) = h^n \prod_j^M (h - z_j )
\prod_{j < k}^M (z_j - z_k )^2 \prod_j^M z_j
\; \; \; ,
\end{equation}

\noindent
with $0 \leq n \leq (N+1)/2$ and $M = (N-1)/2$.  The effect of the
spin-exchange part of the Hamiltonian is the same as for the spinon.
Taylor expanding as usual we obtain

\begin{displaymath}
\sum_{\alpha \neq \beta}^N [ \frac{S_\alpha^+ S_\beta^-}
{| z_\alpha - z_\beta |^2} \Psi_n ]
( z_1 , ... , z_M | h ) = \biggl\{
\biggl[ \frac{1 - N^2}{24} - \sum_{j \neq k}^M
\frac{1}{|z_j -z_k|^2} \biggr] \; h^n
\end{displaymath}

\begin{equation}
+ \frac{N-3}{2} h^{n+1} \frac{\partial}{\partial h} - h^{n+2}
\frac{\partial^2}{\partial h^2} \biggr\} \biggl\{
\frac{\Psi_n (z_1, ... ,z_M | h)}{h^n} \biggr\}
\; \; \; .
\end{equation}

\noindent
The charge-exchange terms also behave similarly.  The operator
$c_{\alpha \downarrow} c_{\beta \downarrow}^\dagger$
gives zero unless the holon resides at $z_\alpha$.  For this case
we have

\begin{displaymath}
\sum_{\beta \neq \alpha}^N [ \frac{P  c_{\alpha \downarrow}
c_{\beta \downarrow}^\dagger P}{|z_\alpha - z_\beta|^2} \Psi_n ]
( z_1 , ... , z_M | z_\alpha ) = \sum_{\beta \neq \alpha}^N \frac{1}
{|z_\alpha - z_\beta |^2} \Psi_n (z_1 , ... , z_M | z_\beta )
\end{displaymath}

\begin{displaymath}
= \sum_{\beta \neq \alpha}^N
\sum_{\ell=0}^{M+1} \frac{z_\beta^n (z_\beta - z_\alpha)^\ell}
{\ell! |z_\beta - z_\alpha|^2}
( \frac{\partial}{\partial z_\alpha} )^\ell \biggl\{
\frac{\Psi_n (z_1, ... , z_M |z_\alpha )}{z_\alpha^n} \biggr\}
\end{displaymath}

\begin{displaymath}
= \biggl\{ \biggl[ \frac{N^2 - 1}{12} + \frac{n(n-N)}{2} \biggr]
z_\alpha^n
\end{displaymath}

\begin{equation}
- \biggl[ \frac{N-1}{2} - n \biggr] z_\alpha^{n+1} \frac{\partial}
{\partial z_\alpha} + \frac{1}{2} z_\alpha^{n+2}
\frac{\partial^2}{\partial z_\alpha^2} \biggr\}
\biggl\{ \frac{\Psi_n (z_1, ... , z_M |z_\alpha )}{z_\alpha^n}
\biggr\}
\; \; \; .
\end{equation}

\noindent
Summing on $\alpha$ we then obtain

\begin{displaymath}
\sum_{\alpha \neq \beta}^N [ \frac{P c_{\alpha
\downarrow} c_{\beta \downarrow}^\dagger P}{|z_\alpha -
z_\beta|} \Psi_n ] ( z_1 , ... , z_M | h ) = \biggl\{ \biggl[
\frac{N^2 - 1}{12} + \frac{n(n-N)}{2} \biggr] h^n
\end{displaymath}

\begin{equation}
- \biggl[ \frac{N-1}{2} - n \biggr] h^{n+1} \frac{\partial}
{\partial h} + \frac{1}{2} h^{n+2} \frac{\partial^2}{\partial h^2}
\biggr\} \biggl\{ \frac{\Psi_n (z_1, ... , z_M | h )}{h^n} \biggr\}
\; \; \; .
\end{equation}

\noindent
For the other charge-exchange channel we use the fact that
$| \Psi_n \! >$ is the same written in the $\downarrow$ coordinates
$\eta_1 , ... , \eta_M$ by virtue of being a singlet.  This gives

\begin{displaymath}
\sum_{\alpha \neq \beta}^N [ \frac{P c_{\alpha
\uparrow} c_{\beta \uparrow}^\dagger  P}{|z_\alpha -
z_\beta|^2} \Psi_n ] ( \eta_1 , ... , \eta_M | h ) = \biggl\{ \biggl[
\frac{N^2 - 1}{12} + \frac{n(n-N)}{2} \biggr] h^n
\end{displaymath}

\begin{equation}
- \frac{1}{2} \biggl[ \frac{N-1}{2} - n \biggr] h^{n+1} \frac{\partial}
{\partial h} + \frac{1}{4} h^{n+2} \frac{\partial^2}{\partial h^2}
\biggr\} \biggl\{ \frac{\Psi_n (\eta_1, ... , \eta_M | h )}{h^n} \biggr\}
\; \; \; .
\end{equation}

\noindent
It remains only to rewrite this expression in terms of the $\uparrow$
coordinates $z_1 , ... , z_M$.  We have

\begin{displaymath}
h^{n+1} \frac{\partial} {\partial h} \biggl\{ \frac{\Psi_n (\eta_1,
... , \eta_M | h )}{h^n} \biggr\}
= \sum_{j}^M \frac{h} {h - \eta_j}  \Psi_n (\eta_1 , ... , \eta_M | h)
\end{displaymath}

\begin{displaymath}
= \biggl\{ \frac{N-1}{2} - \sum_{j}^M \frac{h}
{h - z_j} \biggr\}  \Psi_n (z_1 , ... , z_M | h )
\end{displaymath}

\begin{equation}
= \biggl\{ (\frac{N-1}{2}) h^n - h^{n+1} \frac{\partial} {\partial h}
\biggl\} \biggl\{ \frac{\Psi_n (z_1, ... , z_M | h )}
{h^n} \biggr\} \; \; \; ,
\end{equation}

\noindent
and

\begin{displaymath}
h^{n+2} \frac{\partial^2}{\partial h^2} \biggl\{ \frac{\Psi_n (\eta_1,
... , \eta_M | h )}{h^n} \biggr\} = \sum_{j \neq k}^M \frac{h^2}
{(h - \eta_j )(h - \eta_k ) }  \Psi_n (\eta_1 , ... , \eta_M | h )
\end{displaymath}

\begin{displaymath}
= \sum_{j}^M \frac{h}{h - \eta_j} \biggl\{ \frac{N-1}{2} - \sum_k^M
\frac{h} {h - z_k} - \frac{h}{h - \eta_j} \biggr\}
\Psi_n (\eta_1 , ... , \eta_M | h )
\end{displaymath}

\begin{displaymath}
= \biggl\{ \biggl[ \frac{N-1}{2} - \sum_k^M \frac{h} {h - z_k}
\biggr]^2 + \frac{(N - 1)(N - 5)}{12}
\end{displaymath}

\begin{displaymath}
+ \sum_j^M (\frac{h}{h - z_j})^2
\biggr\} \Psi_n (z_1 , ... , z_M | h )
\end{displaymath}

\begin{displaymath}
= \biggl\{ \biggl[ \frac{(N - 1)(N-2)}{3} + 2 \sum_j^M
( \frac{h}{h - z_j})^2 \biggr] h^n
\end{displaymath}

\begin{equation}
- (N - 1) h^{n+1} \frac{\partial}{\partial h} + h^{n+2}
\frac{\partial^2}{\partial h^2} \biggr\}
\biggl\{ \frac{\Psi_n (z_1, ... , z_M | h )}
{h^n} \biggr\} \; \; \; ,
\end{equation}

\noindent
which gives

\begin{displaymath}
\sum_{\alpha \neq \beta}^N [ \frac{P  c_{\alpha
\uparrow} c_{\beta \uparrow}^\dagger  P}{|z_\alpha -
z_\beta|^2 } \Psi_n ] ( z_1 , ... , z_M | h )
= \biggl\{ \frac{n(n-1)}{2} h^n
+ \sum_j^M \frac{h^{n+2}}{(h - z_j)^2}
\end{displaymath}

\begin{equation}
- n \; h^{n+1}
\frac{\partial}{\partial h}
+ \frac{1}{2} h^{n+2} \frac{ \partial^2}{\partial
h^2} \biggr\} \biggl\{
\frac{ \Psi_n (z_1,...,z_M | h)}{h^n} \biggr\} \; \; \; .
\end{equation}

\noindent
For the Ising and electrostatic energies we have

\begin{displaymath}
\sum_{\alpha \neq \beta}^N [ \frac{S_\alpha^z S_\beta^z}
{| z_\alpha - z_\beta |^2} \Psi_n ]
( z_1 , ... , z_M | h )
\end{displaymath}

\begin{equation}
= \biggl\{ \sum_{j \neq k}^M \frac{1}{|z_j - z_k |^2}
+ \sum_j^M \frac{1}{|h - z_j |^2} - \frac{N(N^2 - 1)}{48} \biggr\}
\Psi_n ( z_1 , ... , z_M | h )
\end{equation}

\noindent
and

\begin{displaymath}
\sum_{\alpha \neq \beta}^N \frac{1} {| z_\alpha - z_\beta |^2}
\biggl[ \frac{1}{2} (n_\alpha + n_\beta) - \frac{1}{4} n_\alpha n_\beta
- \frac{3}{4} \biggr]
\Psi_n ( z_1 , ... , z_M | h )
\end{displaymath}

\begin{equation}
= \frac{1 - N^2}{24} \Psi_n ( z_1 , ... , z_M | h )
\; \; \; .
\end{equation}

\noindent
Adding these contributions together we obtain

\begin{displaymath}
\sum_{\alpha \neq \beta}^N \frac{1}{|z_\alpha - z_\beta |^2} \;
P \biggl[ S_\alpha^+ S_\beta^- + S_\alpha^z S_\beta^z + \sum_s
c_{\alpha s} c_{\beta s}^\dagger + \frac{1}{2}(n_\alpha + n_\beta)
\end{displaymath}

\begin{displaymath}
- \frac{1}{4} n_\alpha n_\beta - \frac{3}{4} \biggr]  P \;
\Psi_n (z_1 , ... , z_M | h )
\end{displaymath}

\begin{equation}
= \biggl[ - \frac{N(N^2 - 1)}{48} + n ( n - \frac{N+1}{2}) \biggr]
\Psi_n (z_1 , ... , z_M | h ) \; \; \; . \; \; \; \Box
\end{equation}

\bigskip

\begin{figure}
\epsfbox{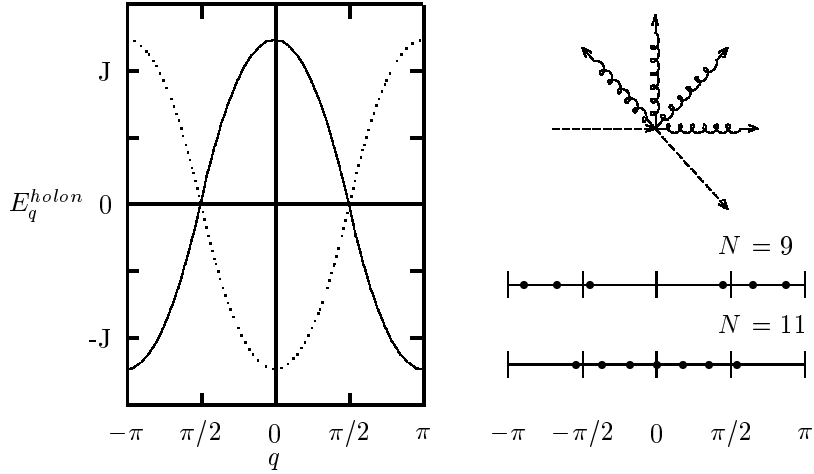}
\caption{Left: Holon dispersion relation defined by Eq. (134).
         Right: Allowed values of $q$ for adjacent odd $N$.
         Only the negative-energy holons are eigenstates of the
         Hamiltonian as positive-energy holons can lose energy
         by spontaneous emission of spinons.}
\end{figure}

\noindent
The propagating holon wavefunction differs from that of the spinon
in being defined for {\it all} values of $m$, including those for which
it is not an eigenstate of the ${\cal H}_{KY}$.  For these other
values of $m$ we observe that

\begin{equation}
0 < \; |< \! \Psi_m | \sum_\alpha^N  P c_{\alpha \downarrow}^\dagger
P | \Psi_m^{holon} \! > |^2 \; < 1
\; \; \; ,
\end{equation}

\noindent
i.e. that a supersymmetric rotation of the spinon, which is an
eigenstate of ${\cal H}_{KY}$, has a nonzero projection onto the
corresponding holon.  The physical meaning of this is that
the holon can lose energy by spontaneous emission of spinons.  Let us
assign to this holon the energy of the exact eigenstate onto
which it projects and write

\begin{equation}
E_m^{\rm holon} = \frac{J}{2} (\frac{2\pi}{N})^2
\left[ \begin{array}{rr}
m (\frac{N+1}{2} + m) & \quad - \frac{N+1}{2} \leq m \leq 0  \\
m (\frac{N-1}{2} - m) & \quad  0 \leq m \leq \frac{N-1}{2}
\end{array} \right] \; \; \; ,
\label{holon1}
\end{equation}

\noindent
or

\begin{equation}
E_q^{holon} = \left[ \begin{array}{rl}
            E_q & \quad |q| < \frac{\pi}{2} \pmod{2 \pi} \\
           -E_q & \quad |q| \geq \frac{\pi}{2} \pmod{2 \pi}
           \end{array} \right] \; \; \; ,
\end{equation}

\noindent
the crystal momentum $q$ and spinon energy $E_q$ are defined as in
Eqs. (\ref{momentum}) and (\ref{dispersion}). This is shown in Fig. 7.
The positive-energy part of the holon band is then unstable because it
has negative curvature and thus does not satisfy the Landau criterion
for spontaneous decay. A negative-energy holon, on the other hand, is
forbidden from decaying by momentum conservation.

\section*{Acknowledgments}

This work was supported primarily by the NSF under grant No.
DMR-9421888.  Additional support was provided by the Center for
Materials Research at Stanford University and by NASA Collaborative
Agreement NCC 2-794. R.~C. and D.~G. would like to thank A.~Devoto,
G.~Morandi, P.~Sodano, A.~Tagliacozzo and V.~Tognetti for giving
them the opportunity to actively participate in the Chia Workshop.
R.~L. would like to express special thanks to A.~Devoto for
creating this excellent school and for his tireless efforts on its
behalf over many years.

\newpage

\appendix

\section{Fourier Sums}

Since the lattice sites $z_\alpha$ are roots of unity we have

\begin{equation}
\prod_\alpha^N (z - z_\alpha ) = z^N - 1 \; \; \; .
\end{equation}

\noindent
Then for $0 < m \leq N$ we have

\begin{figure}
\epsfbox{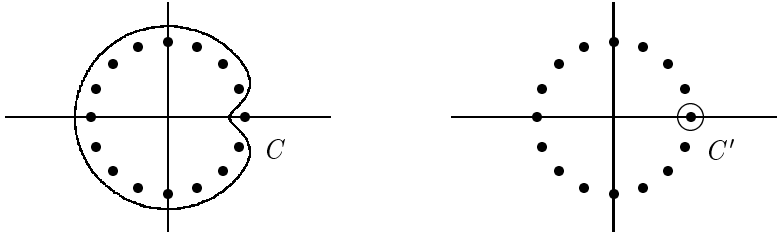}
\caption{Contours used in Eqs. (136) and (137).}
\end{figure}

\begin{displaymath}
\sum_{\alpha = 1}^{N-1} \frac{z_\alpha^m}{z_\alpha - 1}
\end{displaymath}

\begin{displaymath}
= \frac{N}{2\pi i} \oint_C \frac{z^{m-1} \; dz}{(z - 1)(z^N - 1)}
= - \frac{N}{2\pi i} \oint_{C'} \frac{z^{m-1} \; dz}{(z - 1)(z^N - 1)}
\end{displaymath}

\begin{displaymath}
= - \frac{N}{2\pi i} \oint \biggl\{ \frac{1 +
( \! \! \begin{array}{c} m-1 \\ 1 \end{array} \! \! ) \; x +
( \! \! \begin{array}{c} m-1 \\ 2 \end{array} \! \! ) \; x^2 + ... }
{ ( \! \! \begin{array}{c} N \\ 1 \end{array} \! \! ) \; +
( \! \! \begin{array}{c} N \\ 2 \end{array} \! \! ) \; x +
( \! \! \begin{array}{c} N \\ 3 \end{array} \! \! )\; x^2 +
... } \biggr\} \; \frac{dx}{x^2}
\end{displaymath}

\begin{equation}
= \frac{N+1}{2} - m
\; \; \; ,
\end{equation}

\noindent
and

\begin{displaymath}
\sum_{\alpha = 1}^{N-1} \frac{z_\alpha^m}{|z_\alpha - 1|^2}
= - \sum_{\alpha = 1}^{N-1} \frac{z_\alpha^{m+1}}{(z_\alpha - 1)^2}
\end{displaymath}

\begin{displaymath}
= - \frac{N}{2\pi i} \oint_C \frac{z^m \; dz}{(z - 1)^2(z^N - 1)}
= \frac{N}{2\pi i} \oint_{C'} \frac{z^m \; dz}{(z - 1)^2(z^N - 1)}
\end{displaymath}

\begin{displaymath}
= \frac{1}{2\pi i} \oint \biggl\{ \frac{1 +
( \! \! \begin{array}{c} m-1 \\ 1 \end{array} \! \! ) \; x +
( \! \! \begin{array}{c} m-1 \\ 2 \end{array} \! \! ) \; x^2 + ... }
{ ( \! \! \begin{array}{c} N \\ 1 \end{array} \! \! ) \; +
( \! \! \begin{array}{c} N \\ 2 \end{array} \! \! ) \; x +
( \! \! \begin{array}{c} N \\ 3 \end{array} \! \! ) \; x^2 +
... } \biggr\} \; \frac{dx}{x^3}
\end{displaymath}

\begin{equation}
= \frac{N^2-1}{12} - \frac{m(N-1)}{2} + \frac{m(m-1)}{2}
\; \; \; .
\end{equation}

\newpage

\section{Problems}

\begin{enumerate}

\item Let $\eta_1 , ... , \eta_{N/2}$ be the sites complementary to
      $z_1 , ... , z_{N/2}$.  Show that

      \begin{equation}
      \prod_{j < k}^{N/2} (z_j - z_k )^2 \prod_{j}^{N/2} z_j =
      \prod_{j < k}^{N/2} (\eta_j - \eta_k )^2 \prod_{j}^{N/2} \eta_j
      \; \; \; .
      \end{equation}

\item Show that for any polynomial $p(z)$ of degree less than N

      \begin{equation}
      \sum_{\beta \neq \alpha} \frac{p(z_\beta )}{z_\alpha - z_\beta }
      = \frac{1}{z_\alpha} \biggl[ N p(0) - \biggl(\frac{N + 1}
      {2}\biggr)p(z_\alpha ) \biggr] + \frac{\partial p}{\partial z}
      (z_\alpha ) \; \; \; .
      \end{equation}

\item Show that for $\alpha \neq \beta \neq \gamma$

      \begin{equation}
      (\vec{S}_\alpha \times \vec{S}_\beta )^\dagger
      = (\vec{S}_\alpha \times \vec{S}_\beta )
      \end{equation}

      \begin{equation}
      \vec{S}_\alpha \cdot ( \vec{S}_\beta \times \vec{S}_\gamma )
      = \vec{S}_\gamma \cdot ( \vec{S}_\alpha \times \vec{S}_\beta )
      = - \vec{S}_\alpha \cdot ( \vec{S}_\gamma \times \vec{S}_\beta )
      \end{equation}

      \begin{equation}
      [ ( \vec{S}_\alpha \cdot \vec{S}_\beta ) ,
      ( \vec{S}_\alpha \times \vec{S}_\beta ) ] = 0
      \end{equation}

      \begin{equation}
      [ ( \vec{S}_\alpha \cdot \vec{S}_\gamma ) ,
      ( \vec{S}_\alpha \times \vec{S}_\beta ) ]
      = i \biggl[ ( \vec{S}_\alpha \cdot \vec{S}_\beta )
      \vec{S}_\gamma - (\vec{S}_\gamma \cdot \vec{S}_\beta )
      \vec{S}_\alpha \biggr] \; \; \; .
      \end{equation}

\item Show that the operator

      \begin{displaymath}
      \vec{R}_\alpha (z)
      \end{displaymath}

      \begin{equation}
      = \sum_{\beta \neq \alpha}
      \biggl[ \biggl(\frac{1-z}{2} \biggr)
      \frac{z_\alpha} {z_\alpha - z_\beta} +
      \biggl( \frac{1+z}{2} \biggr) \frac{z_\beta} {z_\alpha - z_\beta}
      \biggr] [ i (\vec{S}_\alpha \times \vec{S}_\beta ) + \vec{S}_\beta ]
      \; \; \; ,
      \end{equation}

      where z is an arbitrary complex number, satisfies

      \begin{equation}
      \vec{R}_\alpha (z) | \Psi \! > = 0 \; \; \; ,
      \end{equation}

      where $| \Psi \! >$ is the Haldane-Shastry ground state, and

      \begin{displaymath}
      \sum_\alpha \vec{R}_\alpha^\dagger (z) \cdot \vec{R}_\alpha (z)
      = \frac{3}{2} \biggl\{ 3 \sum_{\alpha \neq \beta}
      \frac{\vec{S}_\alpha \cdot \vec{S}_\beta}
      {|z_\alpha - z_\beta|^2} + \frac{N(N^2 + 5)}{16}
      \end{displaymath}

      \begin{equation}
      + \biggl[ \frac{N-1}{4} \biggl( | z |^2 - 1 \biggr) - \frac{1}
      {2} \biggr] S^2 - \biggl( \frac{z + z^*}{2} \biggr) \vec{S}
      \cdot \vec{\Lambda} \biggr\}
      \; \; \; .
      \end{equation}

\item Show that the ground state wavefunction for a noninteracting
      fermi sea on this lattice is

      \begin{equation}
      \Phi(z_1 , ... , z_{N/2}, \eta_1 , ... , \eta_{N/2})
      = \prod_j^{N/2} (z_j \eta_j)^{-N/2}
      \prod_{j \leq k}^{N/2} (z_j - z_k) (\eta_j - \eta_k) \; \; \; .
      \end{equation}

      Then show that the Haldane-Shastry ground state is the
      Gutzwiller projection of $| \Phi \! >$, i.e. that
      $| \Psi_{HS} \! > = P \; | \Phi \! >$,
      where $P$ is defined as in Eq. (\ref{gutz}). Hint: If p
      denotes a permutation of $N/2$ things and sgn$(p)$ is its
      sign, then

      \begin{equation}
      \sum_p^{(N/2)!} {\rm sgn}(p) \;
      z_{p(1)}^0 \times ... \times z_{p(N/2)}^{N/2 -1} =
      \prod_{j<k}^{N/2} (z_j - z_k) \; \; \; .
      \end{equation}

\item For lattice site $z_\alpha$ different from 1 show that

      \begin{displaymath}
      \frac{2}{J} ( \frac{N}{2\pi})^2 \sum_{m = -(N+1)/2}^{(N-1)/2}
      E_m^{\rm holon} z_\alpha^m = \frac{z_\alpha}{(1 - z_\alpha)^2}
      \biggl\{ \frac{z_\alpha^{(N+1)/2} - z_\alpha^{(N-1)/2}}{2}
      \end{displaymath}

      \begin{equation}
      + \biggl[ \frac{1 + z_\alpha}{1 - z_\alpha} + \frac{1}{2} \biggr]
      \biggl[z_\alpha^{(N+1)/2} +  z_\alpha^{(N-1)/2} - 2 \biggr]
      \biggr\} \; \; \; ,
      \end{equation}

      where $E_m^{\rm holon}$ is given by Eq. (\ref{holon1}).  Then
      use this result to show that

      \begin{displaymath}
      < \! \psi_\alpha | \psi_\beta \! > = \sum_{z_1} ...
      \sum_{z_M} \psi_\alpha^*(z_1 , ... z_M) \psi_\beta(z_1 , ... ,
      z_M)
      \end{displaymath}

      \begin{equation}
      = \frac{ 2< \! \psi_\alpha | \psi_\alpha \! >}{N} \biggl\{
      \frac{1 - (z_\alpha^* z_\beta)^{(N+1)/2}}{1 - z_\alpha^* z_\beta}
      - \frac{ 1 + (z_\alpha^* z_\beta)^{(N-1)/2}}{4}
      \biggr\} \; \; \; ,
      \end{equation}

      where $| \psi_\alpha \! >$ and $| \psi_\beta \! >$ are localized
      spinon wavefunctions defined per Eq. (\ref{spinon1}).  Hint:
      The fourier sum measures $< \psi_\alpha |
      c_{\alpha \downarrow}^\dagger {\cal H}_{KY} c_{\beta \downarrow}
      | \psi_\beta \! >$.

\item The amplitudes $a_\ell$ defined by Eq. (\ref{fmn}) are the
      right eigenvectors of the non-hermitian matrix of Eq.
      (\ref{hfmn}).  Show that the left eigenvectors of this matrix are

      \begin{equation}
      b_\ell = \frac{1}{2\ell [\ell + m - n - 1/2]}
      \sum_{k=\ell+1}^{0} (m - n + 2 k) b_k
      \; \; \; \; \; \; \; \; \; \;
      (b_0 = 0) \; \; \; ,
      \end{equation}

      for $\ell \leq 0$.  The size of these ``adjoint'' coefficients may
      be judged from Fig. 5, where we have plotted the function
      $\bar{f}_{mn}(z) = \sum_\ell b_\ell z^\ell$.  Then show that

      \begin{equation}
      \sum_\alpha^N z_\alpha^k \; S_\alpha^- | \psi \! >
      = \sum_{j=0}^{k/2} c_j \; | \psi_{k-j , j} \! >
      \; \; \; ,
      \end{equation}

      where $| \psi \! >$ is the Haldane-Shastry ground state
      $| \psi_{mn} \! >$ is the 2-spinon eigenstate defined by Eq.
      (\ref{triplet}), and $c_j$ are a set of coefficients.  Obtain
      an expression these in terms of the $b_\ell$.

\item Let $A_\beta$ and $B_\beta$ be any quantum operators and let

      \begin{displaymath}
      a(z) = \biggl[ \frac{z^{N/2} - z^{-N/2}}{2N}\biggr]
      \sum_\beta^N \biggl( \frac{z + z_\beta}{z - z_\beta} \biggr)
      A_\beta
      \end{displaymath}

      \begin{equation}
      b(z) = \biggl[ \frac{z^{N/2} - z^{-N/2}}{2N}\biggr]
      \sum_\beta^N \biggl( \frac{z + z_\beta}{z - z_\beta} \biggr)
      B_\beta \; \; \; .
      \end{equation}

      Show that

      \begin{equation}
      \frac{N}{2 \pi i} \oint a^\dagger (z) b(z) \; \frac{dz}{z}
      = \sum_\beta^N A_\beta^\dagger B_\beta \; \; \; .
      \end{equation}

\item Show that the operator $\sigma (z)$ defined by Eq. (\ref{sigma})
      is hermitian whenever $|z| = 1$.

\item Let N be odd and let $z_1 , ... , z_M$ and $\eta_1 , ... , \eta_M$
      with $M = (N-1)/2$ be distinct lattice sites.  Show that

      \begin{equation}
      \sum_{j \neq k}^M \frac{1}{|\eta_j - \eta_k |^2} =
      \sum_{j \neq k}^M \frac{1}{|z_j - z_k |^2} + \sum_{j=1}^M
      \frac{1}{|h - z_j |^2} - \frac{N^2 - 1}{12} \; \; \; ,
      \end{equation}

      where $h$ denotes the one site not occupied by $z_j$ or $\eta_j$.

\end{enumerate}

\end{document}